%% file: HIN-18-001_temp.tex
\pdfoutput=1

\documentclass[11pt,twoside,a4paper,cmspaper,final,collab]{cms-tdr}
\def\svnVersion{53a4d1f-D}\def\svnDate{2019/10/09}\def\cmsCernNoTag{CERN-EP-2018-345}\def\cmsCernDate{\today}\def\cmsMessage{"Published in Physical Review C as \href{http://dx.doi.org/10.1103/PhysRevC.100.044902}{\doi{10.1103/PhysRevC.100.044902}.}"}

\begin{document}\cmsNoteHeader{HIN-18-001}

\hyphenation{had-ron-i-za-tion}
\hyphenation{cal-or-i-me-ter}
\hyphenation{de-vices}
\RCS$HeadURL: svn+ssh://mstojano@svn.cern.ch/reps/tdr2/papers/HIN-18-001/trunk/HIN-18-001.tex $
\RCS$Id: HIN-18-001.tex 487029 2019-01-23 16:36:28Z alverson $

\newlength\cmsFigWidth
\ifthenelse{\boolean{cms@external}}{\setlength\cmsFigWidth{0.85\columnwidth}}{\setlength\cmsFigWidth{0.4\textwidth}}
\ifthenelse{\boolean{cms@external}}{\providecommand{\cmsLeft}{upper\xspace}}{\providecommand{\cmsLeft}{left\xspace}}
\ifthenelse{\boolean{cms@external}}{\providecommand{\cmsRight}{lower\xspace}}{\providecommand{\cmsRight}{right\xspace}}

\newcommand{\PbPb}{\ensuremath{\text{PbPb}}\xspace}
\newcommand{\XeXe}{\ensuremath{\text{XeXe}}\xspace}
\newcommand {\pPb}  {\ensuremath{\Pp\text{Pb}}\xspace}

\cmsNoteHeader{HIN-18-001}

\title{Charged-particle angular correlations in \XeXe collisions at \texorpdfstring{$\sqrtsNN=5.44\TeV$}{sqrt(s[NN])=5.44 TeV}}

\date{\today}

\abstract{
Azimuthal correlations of charged particles in xenon-xenon  collisions at a center-of-mass energy per nucleon
pair of $ \sqrtsNN =  5.44\TeV$ are studied. The data were collected by the CMS experiment at the
LHC with a total integrated luminosity of $3.42\mubinv$.  The collective motion of the system
formed in the collision is parameterized by a Fourier expansion of the azimuthal particle density distribution.
The azimuthal anisotropy coefficients $v_{2}$, $v_{3}$, and $v_{4}$ are obtained by the scalar-product,
two-particle correlation, and  multiparticle correlation methods.
Within a hydrodynamic picture, these methods have different sensitivities to  non-collective and fluctuation effects.
The dependence of the Fourier coefficients on the size of the colliding system is explored by comparing the xenon-xenon results with equivalent lead-lead data.
Model calculations that include initial-state fluctuation effects are also compared to the experimental results.
The observed angular correlations provide new constraints on the hydrodynamic description of heavy ion collisions.
}

\hypersetup{%
pdfauthor={CMS Collaboration},%
pdftitle={Anisotropy flow in XeXe collisions at 5.44 TeV},%
pdfsubject={CMS},%
pdfkeywords={CMS, heavy ions, QGP, flow, system size}}

\maketitle

\section{Introduction \label{sec:intro}}

At sufficiently high temperatures or densities, lattice quantum chromodynamics predicts a
transition from ordinary hadronic matter to a state of deconfined quarks and
gluons, the so-called quark gluon plasma (QGP) (see, e.g., Ref.~\cite{Karsch:lqcd}).
The QGP state can be reached through relativistic heavy ion collisions, where the collective behavior of the
created medium manifests itself in azimuthal correlations among the emitted particles.  These correlations have been studied in
 gold-gold collisions at the BNL RHIC~\cite{BRAHAMS:2005,PHOBOS:2005,STAR:2005,PHENIX:2005},
lead-lead (\PbPb) collisions at the CERN LHC ~\cite{ALICE:2018flow,ATLAS:2018flow,CMSprc:2014ho},
as well as in collisions involving lighter nuclei, such as the copper-copper system studied at RHIC~\cite{STAR:2010,PHENIX:2007}.
More recently, collective behavior similar to that observed in collisions of heavy nuclei
has also been found in high-multiplicity events produced in the proton-lead (\pPb)
system, and in proton-proton ($\Pp\Pp$) collisions~\cite{ALICE:2014pPb,Khachatryan:2015waa,CMSplb:2017pp,ATLAS:2017pp}.
The results from these small systems raise the question as to how the size of the colliding system affects the
onset of QGP formation. Measurements from xenon-xenon (\XeXe) collisions, as presented here,
bridge the gap between the  small ($\Pp\Pp$ and \pPb)
and large (\PbPb) systems previously studied at LHC energies.

Anisotropic flow can be characterized by a Fourier expansion~\cite{Ollitrault:1993,Voloshin:1994,Poskanzer:1998yz},
\begin{equation}
\label{FourierDec}
\dfrac{2\pi}{N}\dfrac{\rd N}{\rd\phi} =  1+\sum\limits_{n=1}^{\infty}2v_{n}\cos[n(\phi-\Psi_n)],
\end{equation}
where ${\rd N}/{\rd\phi}$ is the azimuthal particle density and  $\phi$  is the particle azimuthal angle with respect to a reference angle  $\Psi_n$.
Different reference angles can be defined. The ``participant plane" angle is the direction of the semiminor axis of the region perpendicular to the beam
direction spanned by the nucleons that undergo a primary interaction.  The ``event-plane" angle is defined by the direction
perpendicular to the beam direction of the maximum outgoing particle
density.  In this paper the measured anisotropies are expressed in terms of the event-plane reference angle.  Averaged over many events, the anisotropies measured with respect to the event plane are expected to be similar to those that would be obtained if it were possible to determine the actual participant plane.

The magnitude of the azimuthal anisotropy is characterized by the  Fourier coefficients $v_{n}$.  The second- and third-order Fourier coefficients are referred to as ``elliptic'' ($v_2$) and ``triangular'' ($v_3$) flow, respectively. The former reflects the lenticular shape of the collision overlap region, as well as initial-state fluctuations in the positions of nucleons at the moment of impact~\cite{Alver:2010prc}. The latter is largely a consequence of fluctuations. While the $v_2$ and $v_3$ harmonics are believed to reflect the initial-state geometry~\cite{Li:2015v7}, for $n\geq 4$ the  flow harmonics are also strongly affected by the dynamics of the system expansion. Hence, studying  both the lower and higher flow harmonics is important for understanding the medium created in heavy ion collisions.

This analysis presents  measurements of the charged-particle collective flow  in \XeXe collisions
at a center-of-mass energy per nucleon pair of $\sqrtsNN=5.44\TeV$. The results are shown as functions of transverse momentum, \pt, for the pseudorapidity region $\abs{\eta} < 2.4$ and for different collision overlap geometries. Spectrum-weighted values with $0.3 < \pt < 3.0\GeVc$, with the efficiency-corrected yield in each $\pt$ interval used as the weight, are also presented.
The Fourier coefficients $v_2$, $v_3$, and $v_4$ are obtained by two-particle correlations ($v_n\{2\}$), the scalar-product method ($v_n\{ {\mathrm{SP}} \}$),
and  multiparticle cumulant analyses ($v_{n}\{m\}$, $m$ = 4, 6, and 8).

Event-by-event fluctuations in the spatial overlap geometry lead to method-dependent differences in the extracted $v_{n}$ values~\cite{Ollitrault:2009v4,Yan:2014flc}.
The  fluctuations cause an increase in the deduced $v_{n}$ values found using two-particle correlations and the scalar-product method,
as compared to the corresponding participant plane value, while the  four-particle cumulant $v_{n}$ results are decreased.
For fluctuations that follow a two-dimensional Gaussian behavior, the flow harmonics based on more than four particles are expected to be the
same as the four-particle correlations results.  Deviations from this common behavior can be used to estimate the higher-order moments
of the fluctuation distribution.   Comparison of flow coefficients measured by different methods probes the initial-state conditions.

The \XeXe values are compared to the results from \PbPb collisions at $\sqrtsNN=5.02\TeV$. The comparison with measurements from different collision systems,
but with similar collision geometry, can give insight to the system size dependence of the anisotropic flow~\cite{ALICE:2018xe}.
Theoretical predictions are compared to the observed system size dependence of the flow harmonics. The results presented here
provide new information on the initial-state geometry and its fluctuations, as well as  the system size dependence of the medium properties.

\section{CMS detector\label{sec:cms}}
The central feature of the CMS apparatus is a superconducting solenoid of 6\unit{m} internal diameter, providing a magnetic field of 3.8\unit{T}.
Within the solenoid volume are a silicon pixel and strip tracker, a lead tungstate crystal electromagnetic calorimeter (ECAL),
and a brass and scintillator hadron calorimeter (HCAL), each composed of a barrel and two endcap sections. Forward calorimeters extend the
pseudorapidity coverage provided by the barrel and endcap detectors. Muons are detected in gas-ionization chambers embedded in the steel
flux-return yoke outside the solenoid. The hadron forward (HF) calorimeter uses steel as an absorber and quartz fibers as the sensitive material.
The two HF calorimeters are located 11.2\unit{m} from the interaction region, one on each end, and together they
provide coverage in the range $3.0 < \abs{\eta} < 5.2$. These calorimeters serve as  luminosity monitors,
are used to establish the event centrality, and provide the  event-plane information for the scalar-product analysis.  The HF calorimeters are azimuthally
subdivided into $20^{\circ}$ modular wedges and further segmented to form $0.175 \times 10^\circ (\Delta\eta \times \Delta\phi$) towers.
The silicon tracker measures charged particles within the pseudorapidity range $\abs{\eta} < 2.5$.
For nonisolated particles of $1 < \pt < 10\GeV/c$ and $\abs{\eta} < 1.4$, the track resolutions are typically 1.5\% in \pt and 25--90 (45--150)\mum
in the transverse (longitudinal) impact parameter \cite{TRK-11-001}. A more detailed description of the CMS detector,
together with a definition of the coordinate system used and the relevant kinematic variables, can be found in Ref.~\cite{Chatrchyan:2008zzk}.
The detailed Monte-Carlo (MC) simulation of the CMS detector response is based on \GEANTfour~\cite{GEANT4}.

\section{Events and track selection \label{sec:evttracksel}}

Results  based on data recorded by CMS during the LHC runs with \XeXe collisions at $\sqrtsNN = 5.44$\TeV in 2017,
with an  integrated luminosity of $3.42\mubinv$, are compared to similar data obtained in 2015 from
\PbPb collisions at $\sqrtsNN = 5.02\TeV$ with an integrated luminosity of $26\mubinv$.
In both  systems, only tracks with $\abs{\eta}<2.4$ and $0.3 < \pt\ < 10.0\GeVc$  are used.

For the \XeXe events, a hardware level (level-1) trigger required at least one tower of the
HF calorimeters to be above a threshold that was fixed to maximize the number of events counted,
while keeping low the noise contamination from electromagnetic scattering and from pileup
(i.e., multiple interactions in the same or neighboring bunch crossings). This trigger also required the presence of
both colliding bunches at the interaction point. The average online pileup fraction was 0.018 per event.
In addition, a high-level trigger was applied that required at least one track in the pixel detector.
Events are further selected offline by requiring at
least 3\GeV of energy being detected in each of three HF calorimeter
towers on either side of the CMS detector and  to have a reconstructed primary vertex,
containing at least two tracks, located within 15\unit{cm} of the nominal
collision point along the beam axis and within 0.2\unit{cm} in the transverse direction.
In addition, contamination from  beam-gas interactions are suppressed by applying a filter where, for each event with more than ten tracks,
at least 25\% of the tracks are required to satisfy a \textit{high purity}~\cite{TRK-11-001}  track quality criteria.
The event selection efficiency is 95\%.
The track reconstruction algorithm is similar to that used for $\Pp\Pp$ collisions~\cite{TRK-11-001}.

For \PbPb collisions, as compared to \XeXe events, there is an additional level-1 trigger requirement
of a coincidence between signals in the HF calorimeters on either side of the CMS detector.
While offline event selection is similar for \PbPb and \XeXe events, for the \PbPb events the
filter to suppress beam-gas interaction is not applied and  pileup contamination is controlled by following
the procedure outlined in~\cite{CMS:2017hpt}.

To ease the computational load for high-multiplicity central \PbPb collisions,
track reconstruction for \PbPb events is done in two iterations.
The first iteration reconstructs tracks from signals
 (``hits") in the silicon pixel and strip detectors compatible with a trajectory of $\pt>0.9\GeVc$.
The second iteration reconstructs tracks compatible with a trajectory of $\pt > 0.2\GeVc$ using solely the pixel detector.
In the final analysis, the first iteration tracks with $\pt > 1.0\GeVc$ are combined with  pixel-detector-only
tracks with $\pt < 2.4\GeVc$, after removing duplicates.

In this paper only tracks from primary charged particles are considered.
For the \XeXe tracks and the \PbPb tracks with both silicon pixel and strip hits,
the impact parameter significance of the tracks with respect to the primary vertex in
both the beam direction ($d_Z$) and the transverse plane ($d_0$)
must be less than three standard deviations, while the relative \pt uncertainty ($\sigma_{\pt}/\pt$) must be below 10\%.
In addition, each track is required to have at least 11 hits in the tracker,
and the chi-square per degree of freedom, associated with fitting the track trajectory,
normalized to the  total  number  of  layers  with  hits  along the trajectory, $\chi^2/\mathrm{dof}/\mathrm{layers}$,  must be less than 0.15.
For the \PbPb pixel-only tracks, it was required that $d_Z$ be less than eight standard deviations and that $\chi^2/\mathrm{dof}/\mathrm{layers}<12$.

\section{Analysis techniques\label{sec:anatech}}

The analysis techniques used in this study are fully described in previous CMS publications.
A two-particle correlation analysis, as discussed in Refs.~\cite{CMSJhep:2011tpc,CMS:2012tpc},
is performed for both the \XeXe and \PbPb data sets.  In addition,  scalar-product and multiparticle cumulant analyses,
as described in Ref.~\cite{Sirunyan:2017igb},
are done for the \XeXe data.

In the two-particle correlation analyses, a charged particle from  one transverse momentum interval is used as a ``trigger" particle,
to be  paired with all of the remaining charged particles from either the same or a different $\pt$ interval,
the ``associated" particles.
For a given trigger particle, the pairing is done in bins of pseudorapidity and azimuthal
angle $(\Delta\eta, \Delta\phi)$.  A similar pairing between the particles randomly chosen from two different events
is done to establish a background distribution.  A Fourier analysis of the azimuthal correlation between the trigger and
associated particles leads to $V_{n\Delta}$ Fourier coefficients, where $n$ is the Fourier order.
If factorization is assumed, the two-particle coefficients can be expressed in terms of single-particle coefficients, with
$V_{n\Delta}(\pt^{\text{trig}},\pt^{\text{assoc}})=v_{n}\{2\}(\pt^{\text{trig}})  v_{n}\{2\}(\pt^{\text{assoc}})$.
The $v_n(\pt^{\text{assoc}})$ term is given by  $\sqrt{V_{n\Delta}(\pt^{\text{assoc}},\pt^{\text{assoc}})}$, thereby allowing $v_n(\pt^{\text{trig}})$ to be
determined.

In order to minimize statistical uncertainties, the associated particles are taken from a wide \pt range with large average anisotropic flow.
In this analysis, $1.0 < \pt^{\text{assoc}} < 3.0 \GeVc$.
To avoid short-range, nonflow correlations,
a pseudorapidity gap of $\abs{\Delta\eta} > 2$ is required for the particle pairs.

The scalar-product event-plane measurements are based on recentered flow $Q$ vectors, defined as:
\begin{linenomath}
\ifthenelse{\boolean{cms@external}}
{
\begin{multline*}
\label{eqn:recenterx}
{\vec Q_n} = \left( \sum\limits_i^M {w_i}\cos \left( {n{\phi _i}} \right) - \left\langle {\sum\limits_i^M {{w_i}\cos \left( {n{\phi _i}} \right)} } \right\rangle \right. ,\\
\left. \sum\limits_i^M {{w_i}\sin \left( {n{\phi _i}} \right) - \left\langle {\sum\limits_i^M {{w_i}\sin \left( {n{\phi _i}} \right)} } \right\rangle }   \right).
\end{multline*}
}
{
\begin{equation*}
\label{eqn:recenterx}
{\vec Q_n} = \left( {\sum\limits_i^M {{w_i}\cos \left( {n{\phi _i}} \right) - \left\langle {\sum\limits_i^M {{w_i}\cos \left( {n{\phi _i}} \right)} } \right\rangle ,\sum\limits_i^M {{w_i}\sin \left( {n{\phi _i}} \right) - \left\langle {\sum\limits_i^M {{w_i}\sin \left( {n{\phi _i}} \right)} } \right\rangle } } } \right).
\end{equation*}
}
\end{linenomath}
\noindent Here, $w_{i}$ is a weight for the $i$th particle emitted at azimuthal angle $\phi_{i}$.
The summations are over the number of particles $M$ within a given (centrality, $\eta$ range, $\pt$ range) analysis bin for a given event.
The averages indicated by the angular brackets are taken over  all particles in all events within each analysis bin.
These averages correspond to the recentering operation and are needed to minimize detector acceptance effects.
If the $Q$ vectors are presented as the corresponding complex scalars, the flow coefficients are given by
\begin{equation}
\label{eq:qsp}
{v_n}\left\{ {{\mathrm{SP}}} \right\} \equiv {\frac{\left\langle {{Q_n}Q_{nA}^*} \right\rangle } {\sqrt {{\frac{\left\langle {{Q_{nA}}Q_{nB}^*} \right\rangle \left\langle {{Q_{nA}}Q_{nC}^*} \right\rangle }  {\left\langle {{Q_{nB}}Q_{nC}^*} \right\rangle }}} }}.
\end{equation}
\noindent The particles of interest are used to obtain the $Q_n$ vector, with unit weighting ($w_i=1$) in the sum.
The subscripts $A$, $B$, and $C$ refer to three separate reference vectors established in different $\eta$ regions.
The product of $Q_n$ with the $Q_{nA}$ reference vector correlates the particles of interest with particles detected in the HF calorimeter (region A).
For the current measurement particles of
interest with $-0.8<\eta<0.0$ ($0.0<\eta<0.8$) and within different \pt ranges are correlated  with HF particles
in the range $3<\eta<5$ ($-5<\eta<-3$).
The  products with $Q$-vectors $B$ and $C$
are used to correct for finite resolution effects.
The  $Q_{nC}$  vector
corresponds to particles detected in the HF calorimeter opposite to that used to define the $Q_{nA}$ vector. The $Q_{nB}$ vector
corresponds to particles measured in the tracker with $\abs{\eta}<0.5$.
Since the $v_n(\pt)$ coefficients increase with $\pt$ up to $\approx 3\GeVc$, the choice of either \pt or $E_{\mathrm{T}}$
weighting results in a better event-plane resolution than with unit weighting. The $Q_{nA}$ and $Q_{nC}$ vectors use $E_{\mathrm{T}}$ weighting, whereas the $Q_{nB}$
vector uses \pt weighting~\cite{Luzum:2012da}.

The $Q$-cumulant method
is used in this analysis to obtain the four- ($v_n\{4\}$), six- ($v_n\{6\}$), and eight- ($v_n\{8\}$) particle $n$th-order harmonic results
by correlating unique combinations of four, six, and eight particles within each event.
The method uses a generic
framework described in Ref.~\cite{Bilandzic:2013kga}. This framework  allows for a track-by-track weighting to correct for the detector acceptance effects.
A wider pseudorapidity range with $\abs{\eta} < 2.4$ is used for the  cumulant method analysis, as compared to the scalar-product method, to reduce
statistical uncertainties.

Results are presented in ranges of collision centrality. The centrality variable is defined as
a fraction of the inelastic hadronic cross section, with 0\% corresponding to full overlap of the two colliding nuclei.
The event centrality is determined offline and is based on the total energy measured in  calorimeters located in the forward pseudorapidity region $3 < \abs{\eta} < 5 $.
The analysis is performed in 11 centrality classes, with intervals ranging from  0--5\% to 60--70\%.
By comparing the \XeXe and \PbPb results in given centrality ranges, similar collision overlap geometries can be achieved, albeit with different numbers of participants.

In comparing the \XeXe and \PbPb results for more peripheral collisions,
it needs to be noted that the \XeXe results can be affected by an
experimental bias introduced by the centrality determination.
Multiplicity fluctuations in the forward region used to determine the event centrality can reduce the centrality resolution.
Monte Carlo studies using the  \HYDJET event generator indicate this bias can be as
large as 5\% in the 50--60\% centrality range and 10\% in the 60--70\% range for the $v_n\{2\}$ coefficients. For the $v_n\{4\}$ coefficients,
the bias is less than 5\% in the 60--70\% centrality range.
For more central events, the bias is found to be negligible.

\section{Systematic uncertainties\label{sec:syst}}

Four different sources of systematic uncertainties are considered.
To study the effect of the track selection on the final results, different track criteria are applied by varying
the limits for the impact parameter significance from 2 to 5, and the relative \pt\ uncertainty from 5\%  to 10\%.
These variations are found to have a 1\% influence on $v_n$ results for peripheral collisions, increasing to 10\%  for the most central collisions at the lowest \pt\ values.
The effect of moving the primary vertex position along the beam axis is studied by comparing the results with events from the vertex position
ranges $\abs{ z_{\mathrm{vtx}} } < 3 \unit{cm}$ and $3 < \abs{ z_{\mathrm{vtx}} } < 15 \unit{cm}$ to the default range of $\abs{ z_{\mathrm{vtx}} } < 15 \unit{cm}$. A 1\% systematic uncertainty is attributed to this source.
The systematic uncertainty resulting from the \XeXe centrality calibration is estimated by varying the event selection criteria.
This uncertainty is largest for the most peripheral centrality bin, where it reaches a value of 3\%.
To explore the sensitivity of the results to the MC simulations on which the efficiency determinations are based,
analyses using the  {\textsc{hydjet 1.9}}~\cite{Lokhtin:2006hmc} event generator are done for generated tracks both before and after the detector effects are taken into account.
The results for the two cases differ by about 2\% for most centrality ranges, but the difference increases to 10\% for the  most central events
and the lowest track transverse momenta, 0.3--0.4\GeVc. The observed differences are included as a %analysis procedure
systematic uncertainty.	
The different uncertainty sources are independent and uncorrelated, therefore the total systematic uncertainty is obtained by combining the individual contributions in quadrature.

\section{Results \label{sec:results}}

\begin{figure*}[bh!]
        \centering
        \includegraphics[width=1\linewidth]{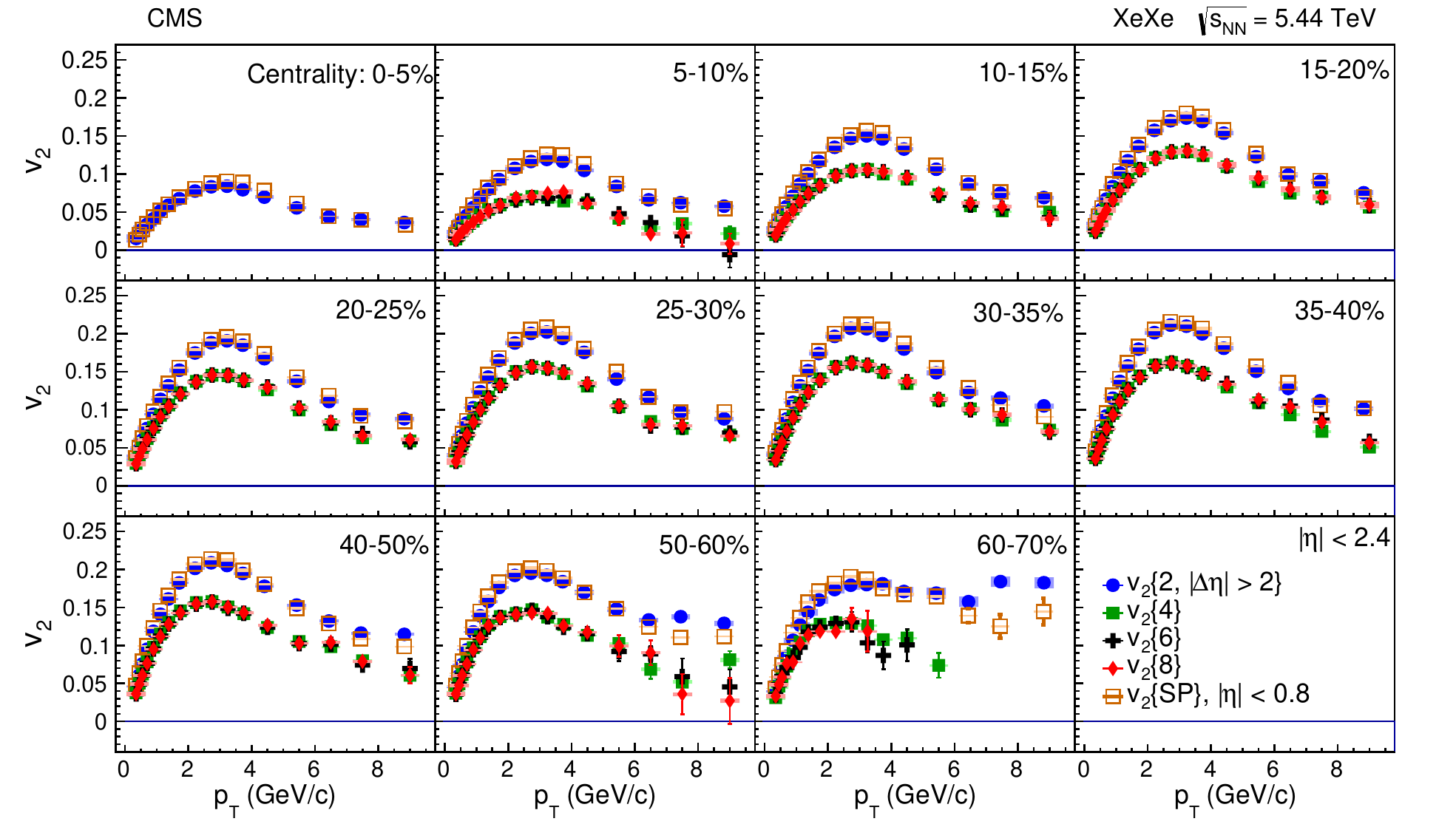}
        \caption{Elliptic-flow coefficients $v_2$ based on different analysis techniques, as functions of transverse momentum and in bins of centrality, from the 5\% most central (top left) to 60--70\% centrality (bottom right). The results for the  two-particle and multiparticle correlations correspond to the range $\abs{\eta} < 2.4$, while the scalar-product results are for $\abs{\eta} < 0.8$. The bars and the shaded boxes represent statistical and systematic uncertainties, respectively.}
        \label{fig:v2_XeXe}
\end{figure*}

Figure~\ref{fig:v2_XeXe} shows the $v_2$ results, as a function of \pt and in 11 centrality bins, as measured with the different techniques.
The two- and multiparticle correlation results are averaged over the pseudorapidity range of $\abs{\eta} < 2.4$, while the scalar-product results are based on tracks with $\abs{\eta} < 0.8$.
The elliptic flow values extracted from two-particle correlations show the same pattern as with the  multiparticle correlations, but with higher magnitudes.
The difference in the results obtained from the two different methods can be largely ascribed to  event-by-event fluctuations of the $v_2$ coefficient~\cite{Ollitrault:2009v4}.
The $v_2$ magnitude increases with \pt, reaching a maximum value of $0.21$ around 3--4\GeVc in the 30--35\% centrality range, and then slowly decreases. The maximum shifts to a lower \pt value as the events become more peripheral.
Whereas $v_2\{\mathrm{SP}\}$ is found to be generally larger than $v_2\{2, \abs{\Delta\eta} > 2\}$, as expected for the narrower range
near midpseudorapidity used for the
scalar-product analysis, the situation switches at higher \pt values for centralities $> 30\%$.  This might reflect a larger nonflow contribution to the
two-particle correlation results.  The pseudorapidity gap of two units used in the two-particle correlation analysis is
less effective in removing non-flow effects, as compared to the gap of three units used for the scalar-product analysis.
In the most peripheral events, the $v_2\{2\}$ distribution becomes almost flat for $\pt > 3.0\GeVc$.
This may be a consequence of nonflow, dijet correlations dominating the results as the system size becomes small.

\begin{figure*}[thbp!]
        \centering
        \includegraphics[width=1\linewidth]{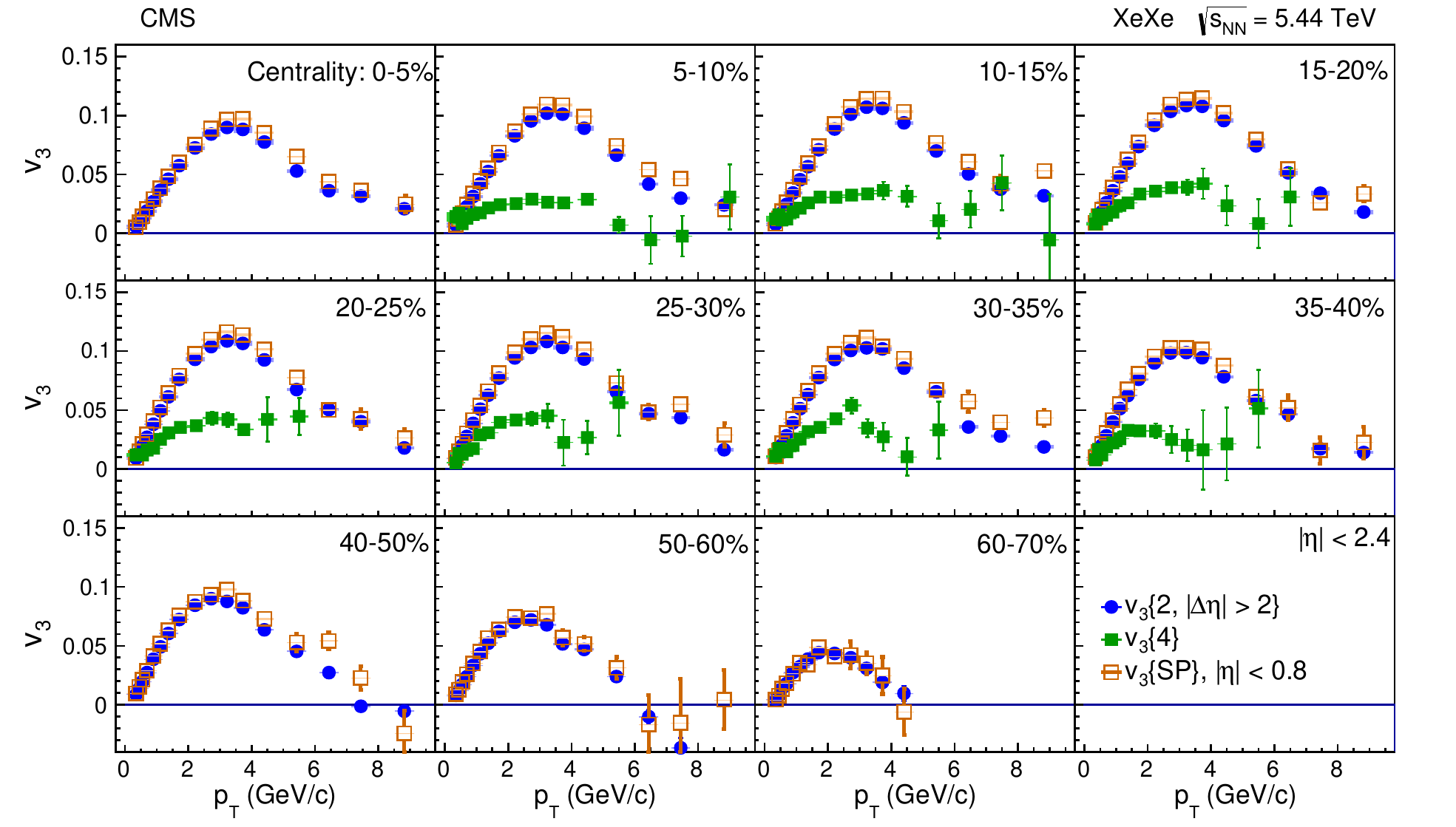}
        \caption{Triangular-flow coefficients $v_3$ based on the different analysis techniques, as functions of transverse momentum and in bins of centrality, from the 5\% most central (top left) to 60--70\% centrality (bottom right). The results for the  two-particle and multiparticle correlations correspond to the range $\abs{\eta} < 2.4$, while the scalar-product results are for $\abs{\eta} < 0.8$. The bars and the shaded boxes represent statistical and systematic uncertainties, respectively.}
        \label{fig:v3_XeXe}
\end{figure*}

Figure~\ref{fig:v3_XeXe} shows the $v_3$ values.
The difference between the two- and four-particle $v_{3}$ values are larger than found for the corresponding $v_{2}$ values,  exceeding a factor of 2.
This suggests a larger fluctuation component to triangular flow as compared to elliptic flow. The difference in amplitude  would be qualitatively expected if the $v_{3}$ correlations were dominated by initial-state fluctuations~\cite{Alver:2010prc}.
For most centralities, the four-particle distributions have no clear peak value and their \pt\ dependence is not as prominent as that found for the two-particle and scalar-product methods.
The $v_3\{m>4\}$ values could not be reliably determined because of their large statistical uncertainties.
The $v_3\{2\}(\pt)$ distribution has a similar shape as found for the  $v_2\{2\}(\pt)$ distribution, but with smaller values that approach zero, or even become negative, at higher \pt values in the most peripheral centrality ranges.

\begin{figure*}[thbp!]
        \centering
        \includegraphics[width=1\linewidth]{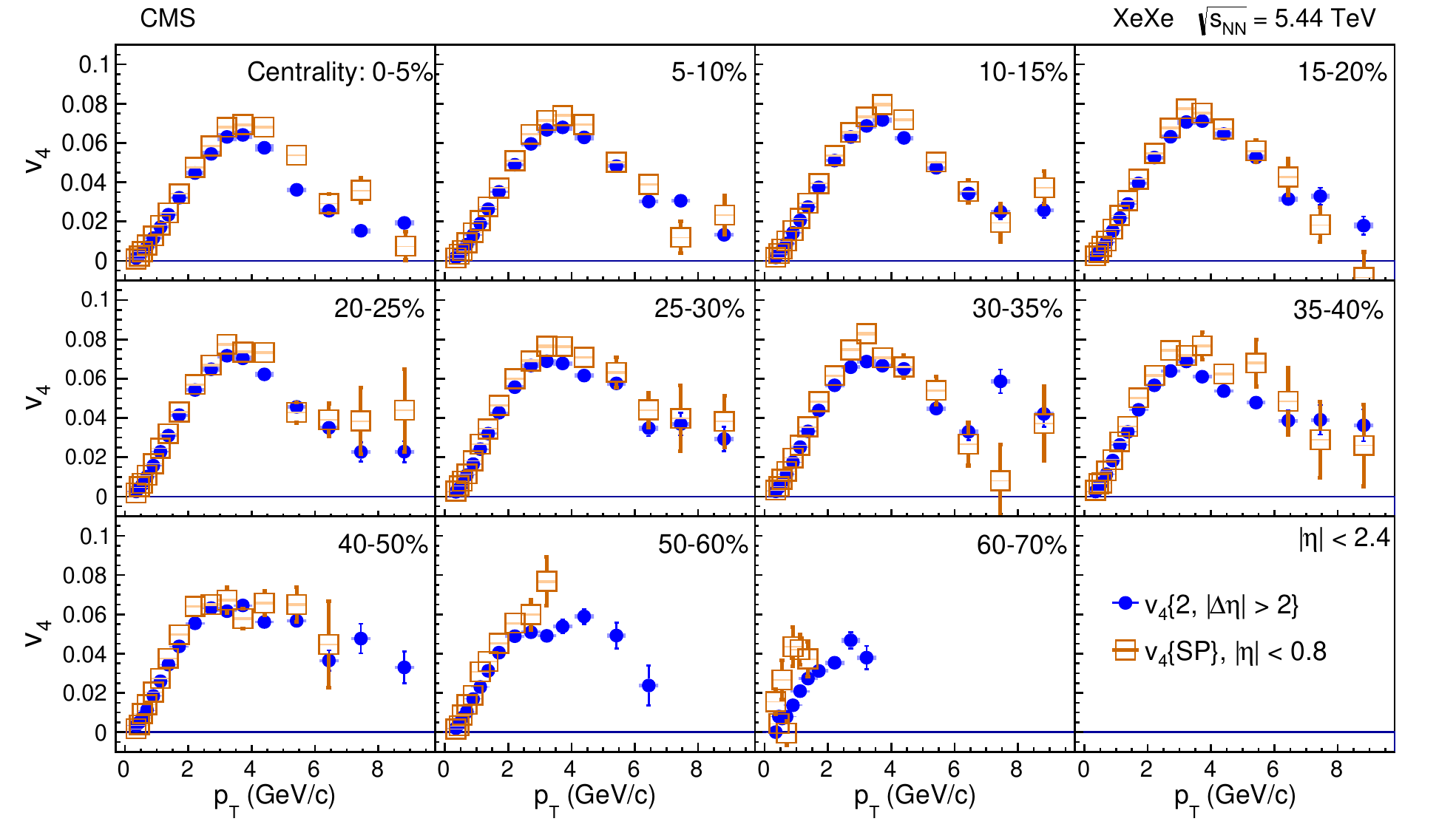}
        \caption{The $v_4$ coefficients, based on the different analysis techniques, as functions of transverse momentum and in bins of centrality, from the 5\% most central (top left) to 60--70\% centrality (bottom right). The results for the  two-particle correlations correspond to the range $\abs{\eta} < 2.4$, while the scalar-product results are for $\abs{\eta} < 0.8$. The bars and the shaded boxes represent statistical and systematic uncertainties, respectively.}
        \label{fig:v4_XeXe}
\end{figure*}

The $v_4$ results from the two-particle correlation and scalar-product  methods are presented in Fig.~\ref{fig:v4_XeXe}.
The $Q$-cumulant results are not shown because of statistical limitations.
The shape of the $v_4(\pt)$ distribution is similar to those for the other measured harmonics.  All three harmonics, with $n$ = 2, 3, and 4,  are found to have
maxima at similar \pt values,  but with the $n$ = 3 and $n$ = 4 harmonics having a reduced centrality dependence as compared to  the $n$ = 2 harmonic.
For all three harmonics, the scalar-product values are systematically larger than the two-particle correlation results.
While fluctuation effects are expected to affect both methods in a similar way, the methods measure flow in different pseudorapidity ranges, which might account for the observed difference.  The similarity of the results  suggests there is only a weak pseudorapidity dependence for all three harmonics.

\begin{figure*}[b!]
        \centering
        \includegraphics[width=1\linewidth]{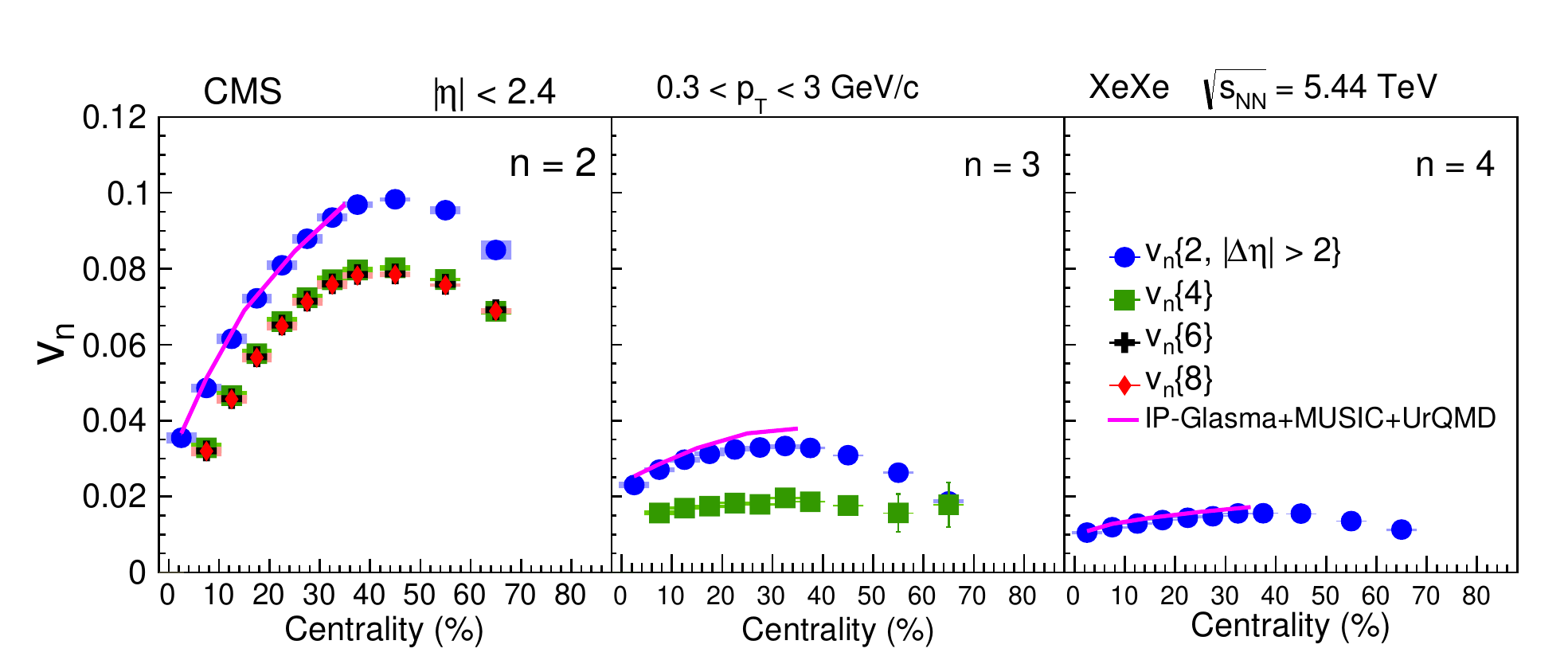}
        \caption{Centrality dependence of the spectrum-weighted $v_2$, $v_3$, and $v_4$ flow harmonics with $0.3 < \pt < 3.0\GeVc$. The $v_2$ results are shown for two-, four-, six-, and eight-particle correlations (left panel). The $v_3$ results are shown for two- and four-particle correlations (middle panel), while the $v_4$ values are presented for two-particle correlations technique, only. The solid curve in each panel is the \textsc{IP-Glasma+music}+UrQMD prediction for $v_n\{2\}$. The shaded boxes represent systematic uncertainties.}
        \label{fig:vn_cent}
\end{figure*}

The spectrum-weighted,
 single-particle anisotropy coefficients, using the two- and multiparticle correlation methods, are presented in Fig.~\ref{fig:vn_cent}. The $v_2$ coefficients show a strong centrality dependence with a maximum value in the 40--50\% centrality bin. The $v_3$ and $v_4$ coefficients have only a weak dependence on centrality.  Results based on  multiparticle cumulants are below the $v_n\{2\}$ values, as expected for the influence of flow fluctuations. The predictions of the \textsc{IP-Glasma+music}+UrQMD model are compared to the experimental $v_n\{2\}$ results.
In this model, initial-state dynamics are described by  impact parameter dependent flowing Glasma gluon fields~\cite{IP:Glasma}.
The subsequent hydrodynamic evolution is calculated with a \textsc{music} simulation~\cite{Schenke:2010ms}, which is a relativistic $(3+1)D$ model that includes shear viscosity (with a shear viscosity over entropy ratio $\eta / s = 0.16$) and a temperature-dependent bulk viscosity over entropy ratio [$\zeta/s(T)$]~\cite{Ryu:2015bv}. The simulation finally switches from a fluid-dynamic description to a
transport description using  the ultrarelativistic quantum molecular dynamics (UrQMD)
 model at the hadronization hypersurface~\cite{UrQMD:2008}. The theoretical calculations are in good agreement with data for the $v_2$ and $v_4$ values. For the $v_3$ coefficient, the calculation gives slightly larger values than observed, with the difference increasing as the size of the nuclear overlap region decreases (i.e., increasing centrality percentage).

\begin{figure*}[thbp!]
        \centering
        \includegraphics[width=1\linewidth]{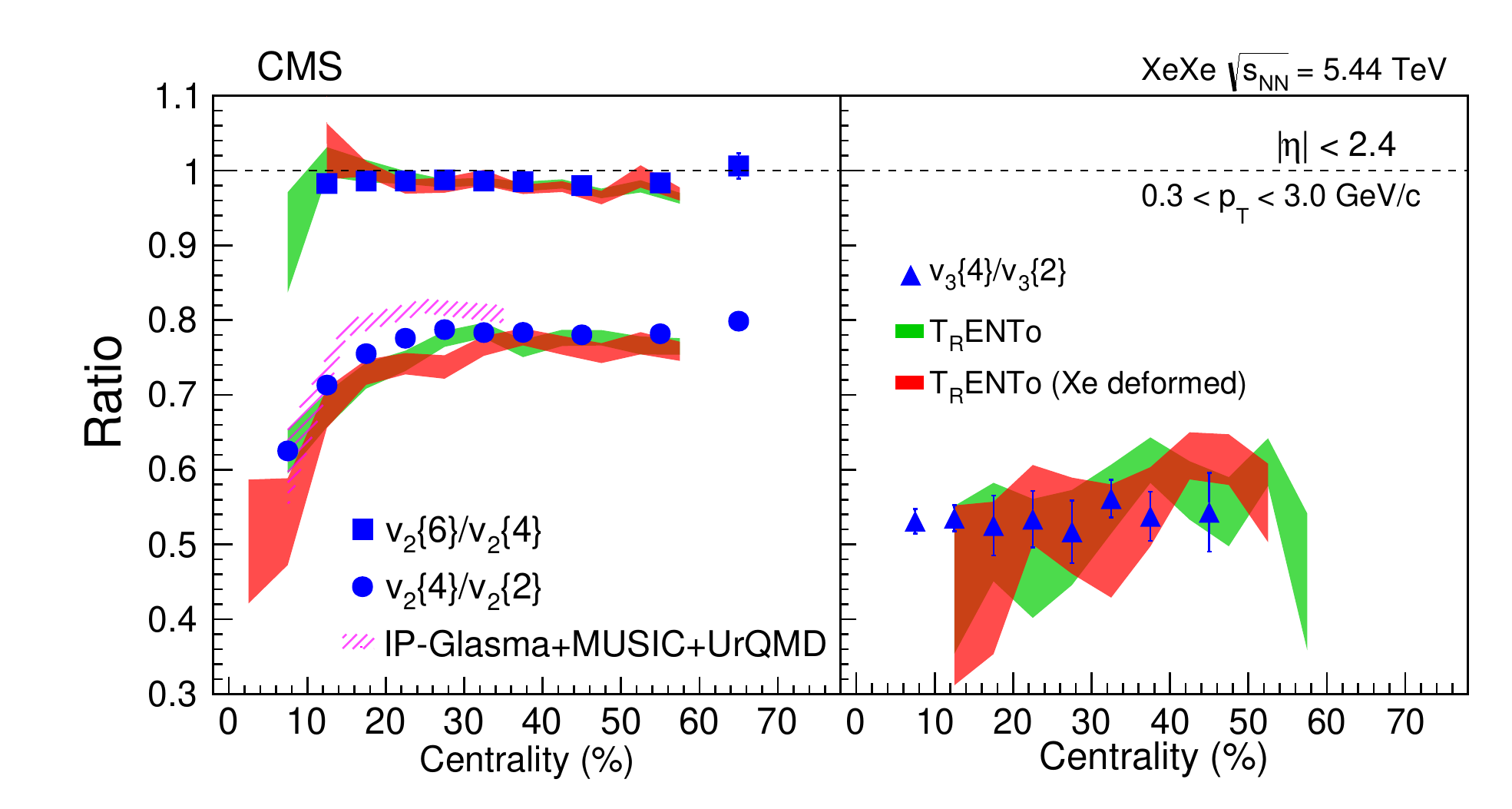}
        \caption{Centrality dependence of  $v_2\{4\}/v_2\{2\}$, $v_2\{6\}/v_2\{4\}$ (left panel) and $v_3\{4\}/v_3\{2\}$ (right panel) ratios. The shaded bands represent the theoretical predictions based on the \textsc{IP-Glasma+music}+UrQMD and the relativistic hydrodynamic model from Ref.~\cite{Giacalone:2017} considering both spherical and deformed xenon nuclei, while the widths of the areas show the statistical uncertainties of the model. The $\mathrm{T_RENTo}$ calculation is done for the \pt range $0.2 < \pt < 5.0\GeVc$.}
        \label{fig:vn_ratio}
\end{figure*}

Figure~\ref{fig:vn_ratio} shows the ratios $v_2\{6\}/v_2\{4\}$,  $v_2\{4\}/v_2\{2\}$, and $v_3\{4\}/v_3\{2\}$.   Theoretical predictions from a hydrodynamic model~\cite{Giacalone:2017} calculation
that uses $\mathrm{T_RENTo}$
 initial conditions~\cite{Moreland:2014tr}  and from the  \textsc{IP-Glasma+music}+UrQMD model are compared to the experimental results.
The former starts the hydrodynamic evolution at a time $\tau = 0.6$ fm/\textit{c}
and has a  shear viscosity to entropy ratio of $\eta / s = $ 0.047.
Xenon is known to be a deformed nucleus with a quadrupole deformation of $\epsilon_2 = 0.15$~\cite{Xe:Shape}.
The $\mathrm{T_RENTo}$ calculations are performed assuming both spherical and nominally deformed xenon nuclei.
The $v_2\{4\}/v_2\{2\}$ ratio shows a strong centrality dependence, with the greatest deviation from unity, with a value of 0.625, corresponding to   5--10\% central events.
The  $v_3\{4\}/v_3\{2\}$ and $v_2\{6\}/v_2\{4\}$ ratios show little, if any, centrality dependence.
The $v_3\{4\}/v_3\{2\}$ has a value close to 0.55 for all centralities, indicating a strong influence of fluctuations on triangular flow~\cite{Ollitrault:2009v4}.
The $v_2\{6\}/v_2\{4\}$ ratio is a few percent below unity and suggests the existence of higher-order
corrections to a near-Gaussian distribution of the event-by-event flow fluctuations~\cite{Giacalone:2017a}.
The \textsc{IP-Glasma+music}+UrQMD and hydrodynamic models give comparable agreement with data for the flow harmonic ratios.
No significant difference is found between the calculations that assume
spherical and deformed Xe nuclear shapes.
This suggests that the fluctuations are not sensitive to the small deformation associated with the nucleus.

\begin{figure*}[thbp!]
        \centering
        \includegraphics[width=1\linewidth]{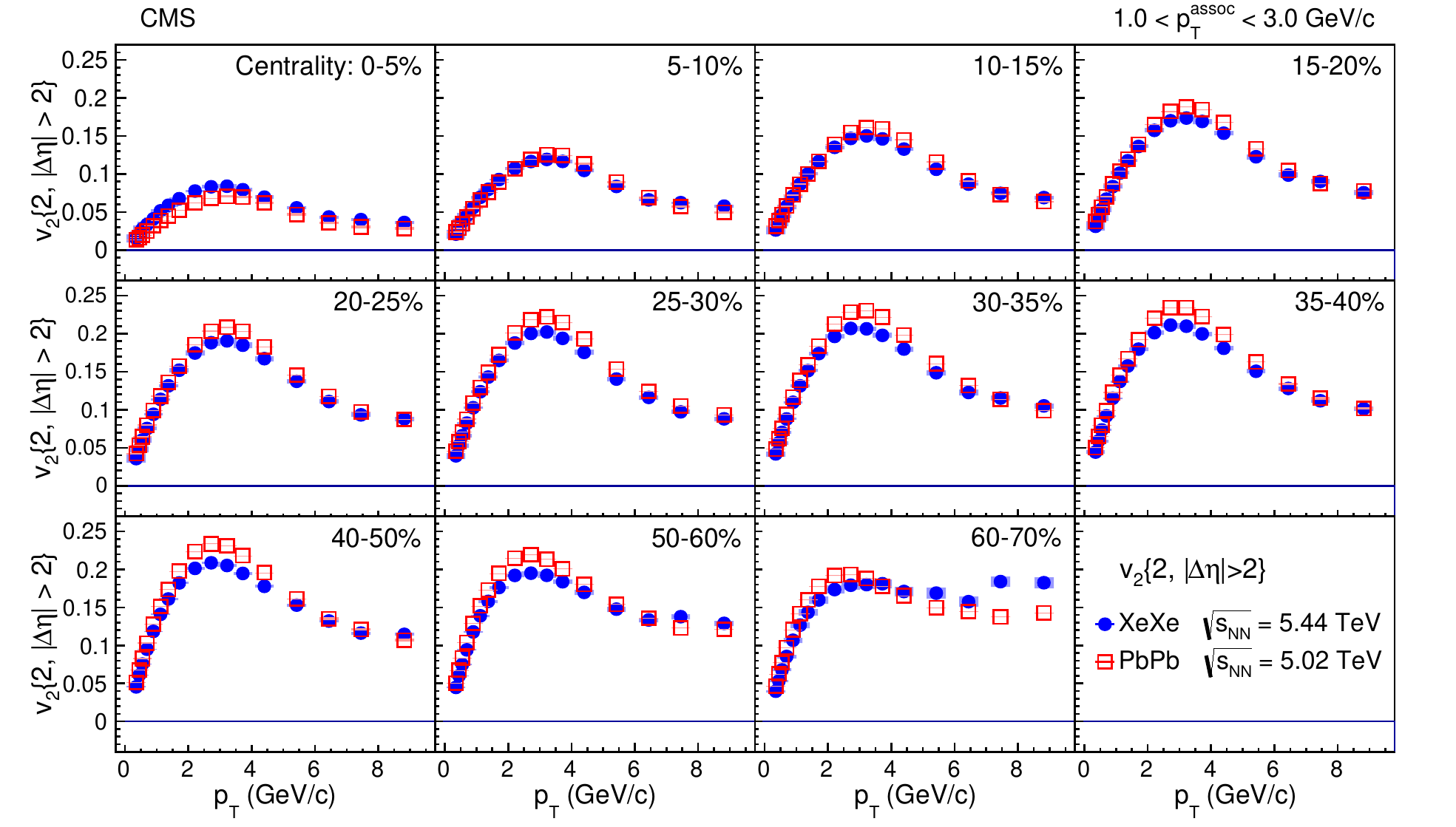}
        \caption{Comparison of the $v_2$ results measured with two-particle correlations from two different systems, \XeXe collisions at $\sqrtsNN=5.44\TeV$ and \PbPb collisions at $5.02\TeV$, shown as a function of \pt\ in eleven centrality bins. The bars (smaller than the marker size) and the shaded boxes represent statistical and systematic uncertainties, respectively.}
        \label{fig:v2_2En}
\end{figure*}

\begin{figure*}[thbp!]
        \centering
        \includegraphics[width=1\linewidth]{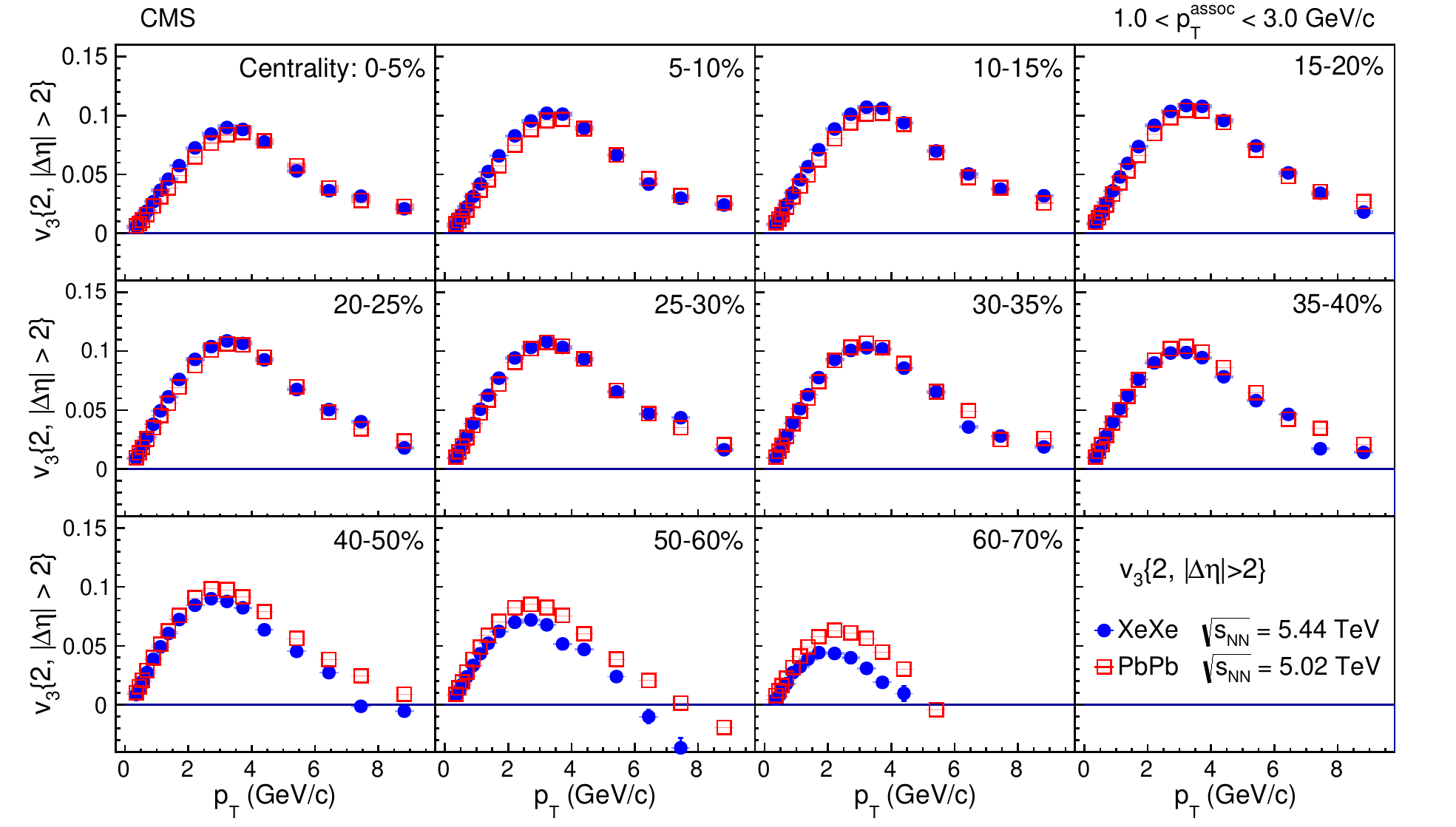}
        \caption{Comparison of the $v_3$ results measured with two-particle correlations from two different systems, \XeXe collisions at $\sqrtsNN=5.44\TeV$ and \PbPb collisions at $5.02\TeV$, shown as a function of \pt\ in 11 centrality bins. The bars (smaller than the marker size) and the shaded boxes represent statistical and systematic uncertainties, respectively.}
        \label{fig:v3_2En}
\end{figure*}

\begin{figure*}[thbp!]
        \centering
        \includegraphics[width=1\linewidth]{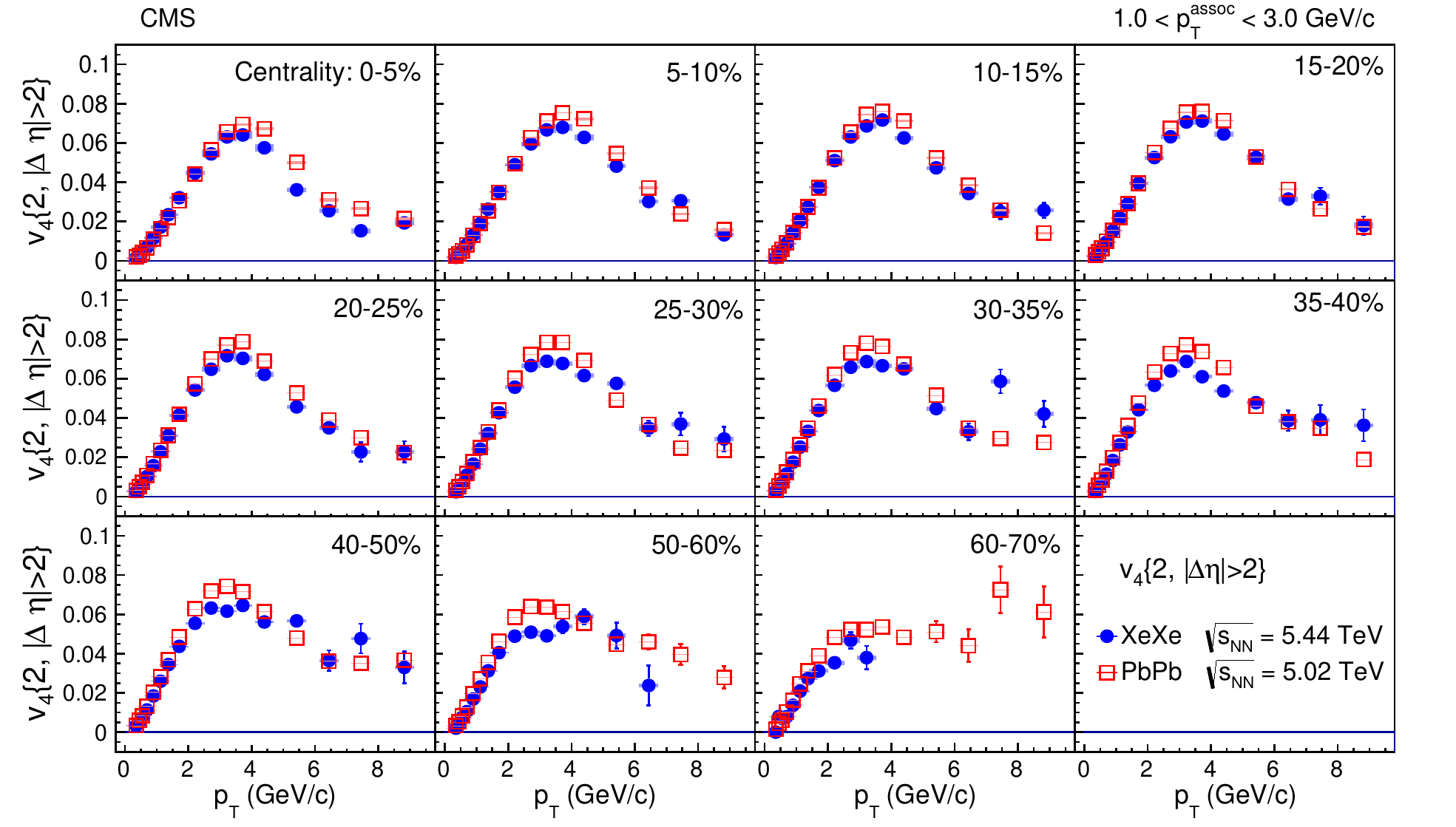}
        \caption{Comparison of the $v_4$ results measured with two-particle correlations from two different systems, \XeXe collisions at $\sqrtsNN=5.44\TeV$ and \PbPb collisions at $5.02\TeV$, shown as a function of \pt\ in 11 centrality bins. The bars and the shaded boxes represent statistical and systematic uncertainties, respectively.}
        \label{fig:v4_2En}
\end{figure*}

The  $v_2$ coefficients obtained by the two-particle correlations technique for \XeXe collisions at $\sqrtsNN =5.44\TeV$ are compared
with corresponding \PbPb data at $5.02\TeV$ as a function of transverse momentum in various centrality bins in Fig.~\ref{fig:v2_2En}.
The $v_2$ values for the two systems show similar dependence on \pt.  However, the  maximum value of the \PbPb elliptic flow coefficient
is found to be greater than the corresponding \XeXe value except in the 0--5\% centrality bin. Since, for the most central collisions,
the participant fluctuations in the
initial-state geometry provide the dominant contribution to the final spacial anisotropy, lower values of $v_2$ in that region are
expected~\cite{Giacalone:2017} for \PbPb collisions because of the larger system size.
The $v_3\{2, \abs{\Delta \eta } > 2\}$ coefficients for the two systems are compared in
Fig.~\ref{fig:v3_2En}. The $v_3$ harmonic is entirely generated by initial participant fluctuations, so slightly larger values are
expected in \XeXe than in \PbPb for central events (e.g., 0--30\% centrality), as observed in the data. However, the $v_3$ harmonic has a larger sensitivity to
transport coefficients (i.e., the shear viscosity) of the created medium, which tends to suppress the azimuthal anisotropy, especially
for systems with a small size. This might explain the  trend of $v_3$  where the system with the larger value
is reversed in the 30--70\% centrality range,
with the larger \PbPb system showing slightly higher $v_3$ values for more peripheral events.
The $v_4\{2, \abs{\Delta \eta}>2\}$ coefficients in \PbPb and \XeXe collisions
are shown in Fig~\ref{fig:v4_2En}.
Higher $v_{4}$ values are found for \PbPb collisions, as compared to the corresponding XeXe collision results, except for
the transverse momentum interval $\pt < 3.0 \GeVc$ in the 5\% most central events.
The ordering of the measured harmonics between the two systems
is consistent with participant fluctuations having a  dominant role in  central collisions, and viscosity effects becoming more important
for mid-central and peripheral collisions.

Since ideal hydrodynamics is scale invariant,  the \XeXe and \PbPb results should have similar behavior~\cite{Giacalone:2017}.
For the same percentage centrality range, the interaction regions of the two colliding systems will have similar average shapes, but will have different size. For example, in the 30--40\% centrality class,
the number of participating nucleons is about 1.6 times higher for the \PbPb collisions.
However, initial-state fluctuations and viscosity corrections can cause scale invariance breaking.
Fluctuations of the initial state are proportional to $A^{-1/2}$, where $A$ is the atomic mass, and, therefore,  one can expect a larger fluctuation component for \XeXe collisions than for \PbPb collisions~\cite{size:fluc}.
However, the influence of the localized fluctuations will decrease with increasing viscosity. The viscosity is thought to be proportional to $A^{-1/3}$~\cite{size:visc}  and is therefore also expected to be larger for \XeXe collisions.
Although the hydrodynamic model simulations do not suggest a large effect on the $v_n\{4\}/v_n\{2\}$ and $v_2\{6\}/v_2\{4\}$
ratios  based on the Xe deformation, this deformation can influence the ratio of the \XeXe and \PbPb results.
The quadrupole deformation of the colliding nuclei is expected to have the greatest influence for the \XeXe $v_2$ values corresponding to the most central collisions~\cite{Giacalone:2017}.

\begin{figure*}[thb!]
        \centering
        \includegraphics[width=1\linewidth]{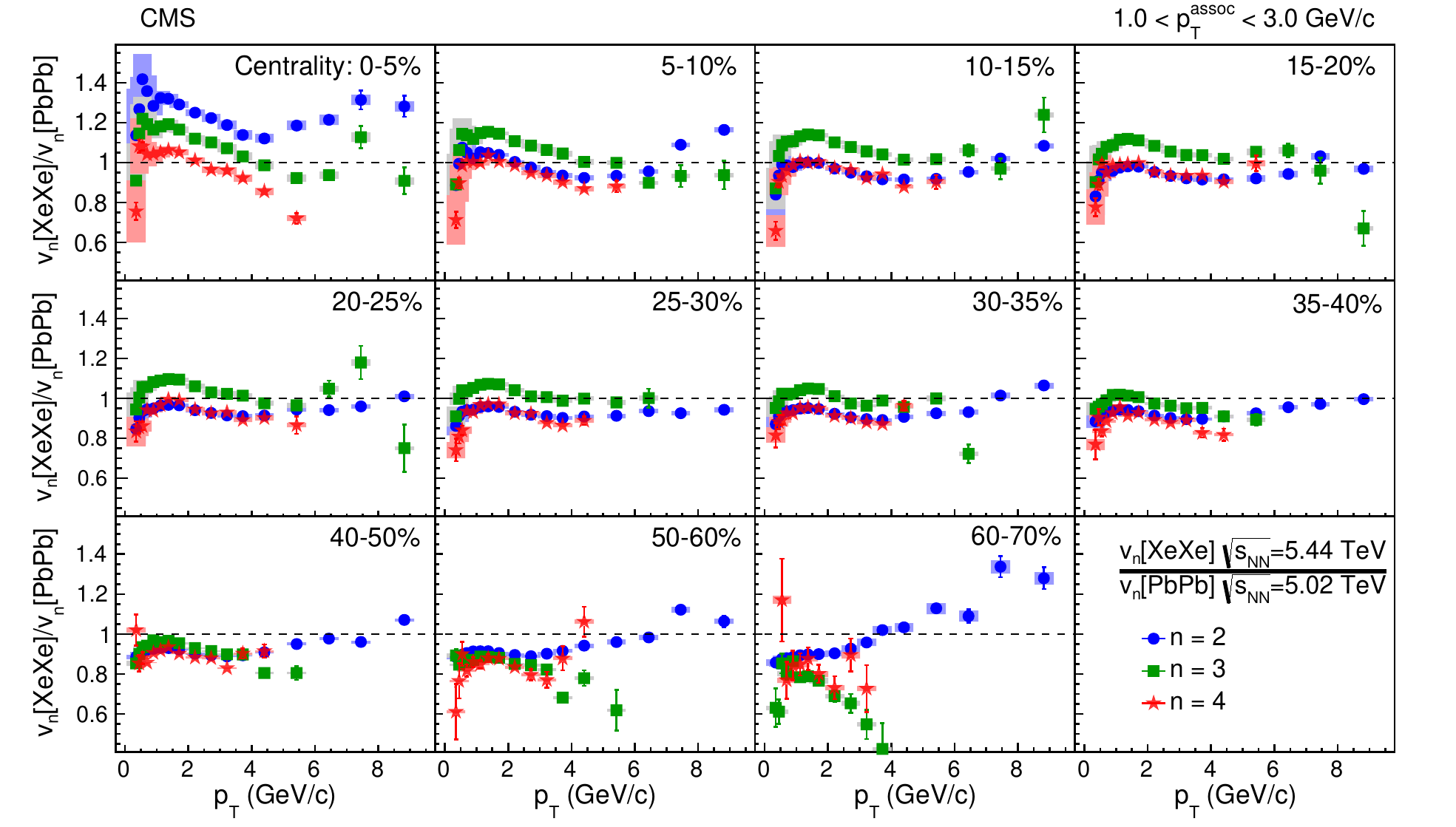}
        \caption{Ratios of the $v_2$, $v_3$, and $v_4$ harmonic coefficients from two-particle correlations in \XeXe and \PbPb collisions as functions of \pt\ in 11 centrality bins. The bars and the shaded boxes represent statistical and systematic uncertainties, respectively.}
        \label{fig:vn_ratio_pT}
\end{figure*}

Figure~\ref{fig:vn_ratio_pT} shows the \pt dependent ratios of \XeXe and \PbPb harmonic coefficients for different centrality ranges.
The ratios reach a maximum value between 1 and 2\GeVc, within the current uncertainties,
 and then decrease up to $\pt \sim 6\GeVc$, at which point they start to increase again.
The increasing trend above  6\GeVc, which is most pronounced for the $v_2$ coefficient, might be a consequence of
back-to-back dijet correlations that can not be fully eliminated with the $\abs{\Delta\eta} > 2$ requirement. This nonflow behavior is increasingly significant
as the system size becomes smaller, with correspondingly smaller particle multiplicities.

\begin{figure*}[thbp!]
        \centering
        \includegraphics[width=1\linewidth]{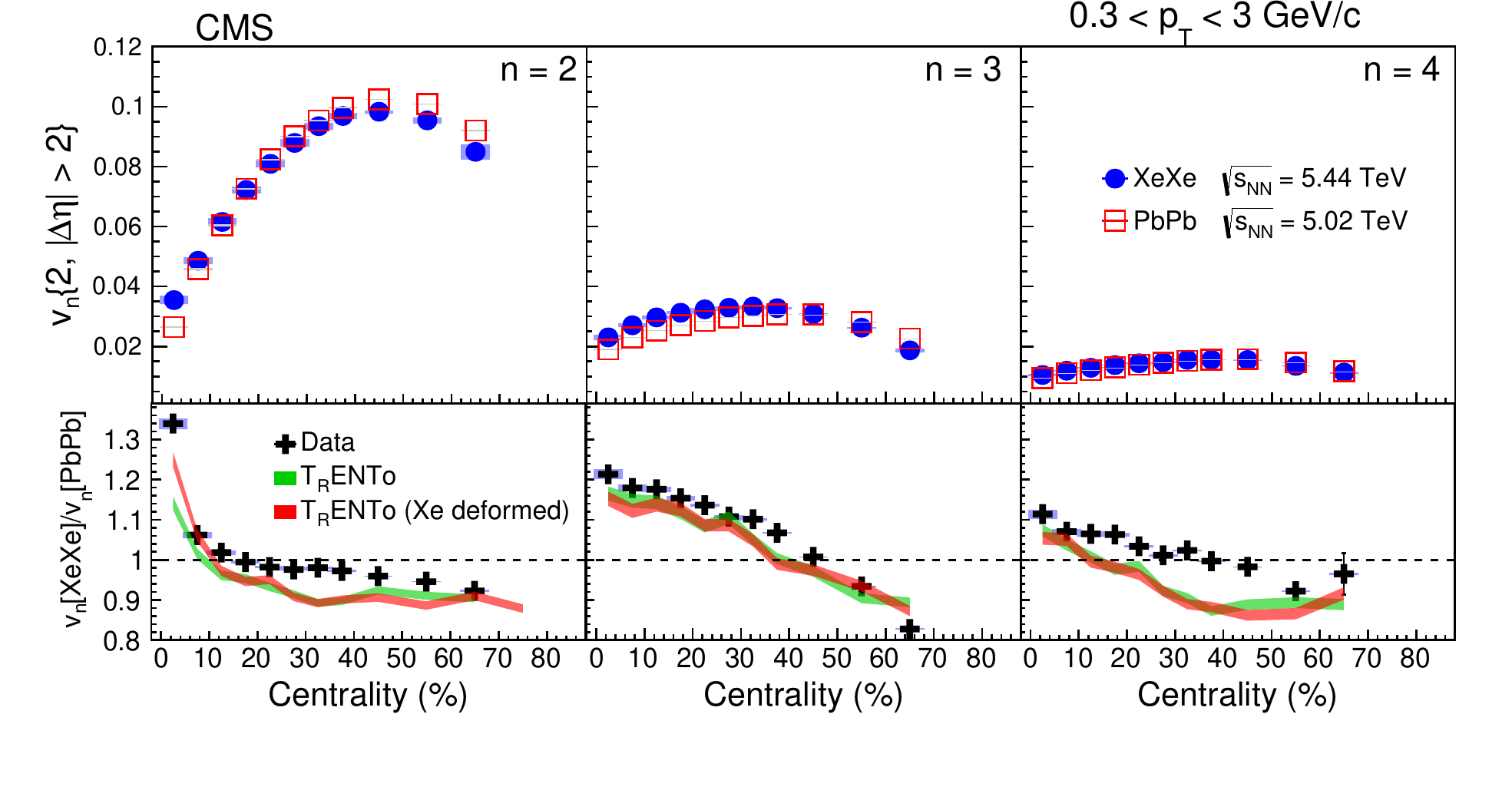}
        \caption{Centrality dependence of the spectrum-weighted $v_2$, $v_3$, and  $v_4$  harmonic coefficients from two-particle correlations method for $0.3 < \pt < 3.0\GeVc$ for  \XeXe collisions at $\sqrtsNN=5.44\TeV$ and \PbPb collisions at $5.02\TeV$.  The lower panels show the ratio of the results for the two systems.   The bars and the shaded boxes represent statistical and systematic uncertainties, respectively. Theoretical predictions from Ref.~\cite{Giacalone:2017} are compared to the data (shaded bands). The model calculation is done for the \pt range $0.2 < \pt < 5.0\GeVc$.}
        \label{fig:vn_centPb}
\end{figure*}

Figure~\ref{fig:vn_centPb}  compares the spectrum-weighted  $v_2$, $v_3$, and $v_4$ values
 with $ 0.3 <  \pt\ < 3.0\GeVc$ for the  \XeXe and \PbPb  systems.
The largest difference between the two systems is found for the $v_2$ coefficients corresponding to the most central events,
where the \XeXe results are larger by a factor of about 1.3.
For centralities above 10\%, the  \PbPb results become higher and the ratio has only a weak centrality dependence.
For the $v_3$ and $v_4$ coefficients,  the ratio $v_n[\text{XeXe}]/v_n[\text{PbPb}]$ decreases with centrality with an almost constant slope.
The relativistic hydrodynamic model calculations of Ref.~\cite{Giacalone:2017} are also shown in Fig.~\ref{fig:vn_centPb}.
Compared to calculations assuming a spherical Xe shape, including the xenon nuclear deformation in hydrodynamic models has little effect on the predicted flow characteristics over the centrality range 10--70\%, as expected. For the most central 0--10\% range, the $v_2[\text{XeXe}]/v_2[\text{PbPb}]$ model ratio shows a greater sensitivity to the xenon nuclear deformation, with the calculation including deformation in better agreement with experiment. For all measured harmonics, the model values lie below the experimental results, with the greatest difference found for the $v_4$ coefficients.

\section{Summary \label{sec:summary}}

In this paper,  the $v_2$, $v_3$, and $v_4$ azimuthal flow harmonics
are shown for xenon-xenon (\XeXe) collisions at
a center-of-mass energy per nucleon pair of $\sqrtsNN = 5.44\TeV$ based on data obtained with the CMS detector.
Three analysis techniques with different sensitivities to flow fluctuations,
including two-particle correlations, the scalar-product method, and the multiparticle cumulant method,
are used to explore the event-by-event fluctuations.
The harmonic coefficients are compared to those found with lead-lead (\PbPb) collisions at $\sqrtsNN = 5.02\TeV$
to explore the effect of the system size.
The magnitude of the $v_2$ coefficients for  \XeXe collisions are larger than those found in \PbPb collisions for the most central collisions.  This is attributed to a larger fluctuation component in the lighter colliding system.
In more peripheral events, the  \PbPb $v_n$ coefficients are consistently larger than those found for \XeXe collisions. This behavior is qualitatively
consistent with expectations from hydrodynamic models.
A clear ordering  $v_2\{2\} > v_2\{4\} \approx v_2\{6\} \approx v_2\{8\}$ is observed for \XeXe collisions, with  $v_2\{6\}$ and $v_2\{4\}$ values differing by 2--3\%.  The $v_3\{4\} / v_3\{2\}$ ratio is found to be significantly smaller than the $v_2\{4\} / v_2\{2\}$ ratio, suggesting a dominant fluctuation component for the $v_3$ harmonic.
Hydrodynamic models that consider the xenon nuclear deformation are able to better describe
the $v_{2}[{\mathrm{XeXe}}]/v_{2}[{\mathrm{PbPb}}]$ ratio in central collisions than those assuming a spherical Xe shape,
although the deformation appears to have little effect on the fluctuation-sensitive ratio of the cumulant orders.
These measurements provide new tests of hydrodynamic models and help
to constrain hydrodynamic descriptions of the nuclear collisions.

\ifthenelse{\boolean{cms@external}}{\clearpage}{}

\begin{acknowledgments}
We congratulate our colleagues in the CERN accelerator departments for the excellent performance of the LHC and thank the technical and administrative staffs at CERN and at other CMS institutes for their contributions to the success of the CMS effort. In addition, we gratefully acknowledge the computing centers and personnel of the Worldwide LHC Computing Grid for delivering so effectively the computing infrastructure essential to our analyses. Finally, we acknowledge the enduring support for the construction and operation of the LHC and the CMS detector provided by the following funding agencies: BMBWF and FWF (Austria); FNRS and FWO (Belgium); CNPq, CAPES, FAPERJ, FAPERGS, and FAPESP (Brazil); MES (Bulgaria); CERN; CAS, MoST, and NSFC (China); COLCIENCIAS (Colombia); MSES and CSF (Croatia); RPF (Cyprus); SENESCYT (Ecuador); MoER, ERC IUT, and ERDF (Estonia); Academy of Finland, MEC, and HIP (Finland); CEA and CNRS/IN2P3 (France); BMBF, DFG, and HGF (Germany); GSRT (Greece); NKFIA (Hungary); DAE and DST (India); IPM (Iran); SFI (Ireland); INFN (Italy); MSIP and NRF (Republic of Korea); MES (Latvia); LAS (Lithuania); MOE and UM (Malaysia); BUAP, CINVESTAV, CONACYT, LNS, SEP, and UASLP-FAI (Mexico); MOS (Montenegro); MBIE (New Zealand); PAEC (Pakistan); MSHE and NSC (Poland); FCT (Portugal); JINR (Dubna); MON, RosAtom, RAS, RFBR, and NRC KI (Russia); MESTD (Serbia); SEIDI, CPAN, PCTI, and FEDER (Spain); MOSTR (Sri Lanka); Swiss Funding Agencies (Switzerland); MST (Taipei); ThEPCenter, IPST, STAR, and NSTDA (Thailand); TUBITAK and TAEK (Turkey); NASU and SFFR (Ukraine); STFC (United Kingdom); DOE and NSF (USA).

\hyphenation{Rachada-pisek} Individuals have received support from the Marie-Curie program and the European Research Council and Horizon 2020 Grant, contract Nos.\ 675440 and 765710 (European Union); the Leventis Foundation; the A.P.\ Sloan Foundation; the Alexander von Humboldt Foundation; the Belgian Federal Science Policy Office; the Fonds pour la Formation \`a la Recherche dans l'Industrie et dans l'Agriculture (FRIA-Belgium); the Agentschap voor Innovatie door Wetenschap en Technologie (IWT-Belgium); the F.R.S.-FNRS and FWO (Belgium) under the ``Excellence of Science -- EOS" -- be.h project n.\ 30820817; the Beijing Municipal Science \& Technology Commission, No. Z181100004218003; the Ministry of Education, Youth and Sports (MEYS) of the Czech Republic; the Lend\"ulet (``Momentum") Program and the J\'anos Bolyai Research Scholarship of the Hungarian Academy of Sciences, the New National Excellence Program \'UNKP, the NKFIA research grants 123842, 123959, 124845, 124850, and 125105 (Hungary); the Council of Science and Industrial Research, India; the HOMING PLUS program of the Foundation for Polish Science, cofinanced from European Union, Regional Development Fund, the Mobility Plus program of the Ministry of Science and Higher Education, the National Science Center (Poland), contracts Harmonia 2014/14/M/ST2/00428, Opus 2014/13/B/ST2/02543, 2014/15/B/ST2/03998, and 2015/19/B/ST2/02861, Sonata-bis 2012/07/E/ST2/01406; the National Priorities Research Program by Qatar National Research Fund; the Programa Estatal de Fomento de la Investigaci{\'o}n Cient{\'i}fica y T{\'e}cnica de Excelencia Mar\'{\i}a de Maeztu, grant MDM-2015-0509 and the Programa Severo Ochoa del Principado de Asturias; the Thalis and Aristeia programs cofinanced by EU-ESF and the Greek NSRF; the Rachadapisek Sompot Fund for Postdoctoral Fellowship, Chulalongkorn University and the Chulalongkorn Academic into Its 2nd Century Project Advancement Project (Thailand); the Welch Foundation, contract C-1845; and the Weston Havens Foundation (USA).
\end{acknowledgments}

\bibliography{auto_generated}

\cleardoublepage \appendix\section{The CMS Collaboration \label{app:collab}}\begin{sloppypar}\hyphenpenalty=5000\widowpenalty=500\clubpenalty=5000\input{HIN-18-001-authorlist.tex}\end{sloppypar}
\end{document}

%% file: HIN-18-001-authorlist.tex
\vskip\cmsinstskip
\textbf{Yerevan Physics Institute, Yerevan, Armenia}\\*[0pt]
A.M.~Sirunyan, A.~Tumasyan
\vskip\cmsinstskip
\textbf{Institut f\"{u}r Hochenergiephysik, Wien, Austria}\\*[0pt]
W.~Adam, F.~Ambrogi, E.~Asilar, T.~Bergauer, J.~Brandstetter, M.~Dragicevic, J.~Er\"{o}, A.~Escalante~Del~Valle, M.~Flechl, R.~Fr\"{u}hwirth\cmsAuthorMark{1}, V.M.~Ghete, J.~Hrubec, M.~Jeitler\cmsAuthorMark{1}, N.~Krammer, I.~Kr\"{a}tschmer, D.~Liko, T.~Madlener, I.~Mikulec, N.~Rad, H.~Rohringer, J.~Schieck\cmsAuthorMark{1}, R.~Sch\"{o}fbeck, M.~Spanring, D.~Spitzbart, A.~Taurok, W.~Waltenberger, J.~Wittmann, C.-E.~Wulz\cmsAuthorMark{1}, M.~Zarucki
% ...
\vskip\cmsinstskip
% ...
1:  Also at Vienna University of Technology, Vienna, Austria\\
% ...


%% file: HIN-18-001_temp.bbl
\providecommand{\href}[2]{#2}\begingroup\raggedright\begin{thebibliography}{10}%
\makeatletter
\providecommand{\hrefCMSnoop }[0]{\@secondoftwo}%
\makeatother
\providecommand{\doi}{\texttt{doi:}\begingroup \urlstyle{tt}\Url}

\bibitem{Karsch:lqcd}
\hrefCMSnoop {}{F.~Karsch, ``Lattice {QCD} at high temperature and density'',}
  in \textit{ Lectures on Quark Matter}, W.~Plessas and L.~Mathelitsch, eds.,
  p.~209.
\newblock Springer Berlin Heidelberg, Berlin, Heidelberg, 2002.
\newblock
  \href{http://www.arXiv.org/abs/hep-lat/0106019}{\texttt{arXiv:hep-lat/0106019}}.
\newblock
\href{http://dx.doi.org/10.1007/3-540-45792-5_6}{\doi{10.1007/3-540-45792-5_6}}.
%%CITATION = ARXIV:HEP-LAT/0106019;%%.

\bibitem{BRAHAMS:2005}
\hrefCMSnoop {}{{BRAHMS} Collaboration, ``{Quark pluon plasma an color glass
  condensate at RHIC? The perspective from the BRAHMS experiment}'',} \textit{
  Nucl. Phys. A} \textbf{ 757} (2005) 1,
  \href{http://dx.doi.org/10.1016/j.nuclphysa.2005.02.130}{\doi{10.1016/j.nuclphysa.2005.02.130}},
\href{http://www.arXiv.org/abs/nucl-ex/0410020}{\texttt{arXiv:nucl-ex/0410020}}.
%%CITATION = NUCL-EX/0410020;%%.

\bibitem{PHOBOS:2005}
\hrefCMSnoop {}{{PHOBOS} Collaboration, ``{The PHOBOS perspective on
  discoveries at RHIC}'',} \textit{ Nucl. Phys. A} \textbf{ 757} (2005) 28,
  \href{http://dx.doi.org/10.1016/j.nuclphysa.2005.03.084}{\doi{10.1016/j.nuclphysa.2005.03.084}},
\href{http://www.arXiv.org/abs/nucl-ex/0410022}{\texttt{arXiv:nucl-ex/0410022}}.
%%CITATION = NUCL-EX/0410022;%%.

\bibitem{STAR:2005}
\hrefCMSnoop {}{{STAR} Collaboration, ``{Experimental and theoretical
  challenges in the search for the quark gluon plasma: The STAR Collaboration's
  critical assessment of the evidence from RHIC collisions}'',} \textit{ Nucl.
  Phys. A} \textbf{ 757} (2005) 102,
  \href{http://dx.doi.org/10.1016/j.nuclphysa.2005.03.085}{\doi{10.1016/j.nuclphysa.2005.03.085}},
\href{http://www.arXiv.org/abs/nucl-ex/0501009}{\texttt{arXiv:nucl-ex/0501009}}.
%%CITATION = NUCL-EX/0501009;%%.

\bibitem{PHENIX:2005}
\hrefCMSnoop {}{{PHENIX} Collaboration, ``{Formation of dense partonic matter
  in relativistic nucleus-nucleus collisions at RHIC: Experimental evaluation
  by the PHENIX collaboration}'',} \textit{ Nucl. Phys. A} \textbf{ 757} (2005)
  184,
  \href{http://dx.doi.org/10.1016/j.nuclphysa.2005.03.086}{\doi{10.1016/j.nuclphysa.2005.03.086}},
\href{http://www.arXiv.org/abs/nucl-ex/0410003}{\texttt{arXiv:nucl-ex/0410003}}.
%%CITATION = NUCL-EX/0410003;%%.

\bibitem{ALICE:2018flow}
\hrefCMSnoop {}{{ALICE Collaboration}, ``{Energy dependence and fluctuations of
  anisotropic flow in Pb-Pb collisions at $\sqrt{s_{_{\rm NN}}} = 5.02$ and
  $2.76$~TeV}'',} \textit{ JHEP} \textbf{ 07} (2018) 103,
  \href{http://dx.doi.org/10.1007/JHEP07(2018)103}{\doi{10.1007/JHEP07(2018)103}},
\href{http://www.arXiv.org/abs/1804.02944}{\texttt{arXiv:1804.02944}}.
%%CITATION = ARXIV:1804.02944;%%.

\bibitem{ATLAS:2018flow}
\hrefCMSnoop {}{{ATLAS Collaboration}, ``{Measurement of longitudinal flow
  decorrelations in Pb+Pb collisions at $\sqrt{s_{_{\rm NN}}} = 2.76$ and
  $5.02$~TeV with the ATLAS detector}'',} \textit{ Eur. Phys. J. C} \textbf{
  78} (2018) 142,
  \href{http://dx.doi.org/10.1140/epjc/s10052-018-5605-7}{\doi{10.1140/epjc/s10052-018-5605-7}},
\href{http://www.arXiv.org/abs/1709.02301}{\texttt{arXiv:1709.02301}}.
%%CITATION = ARXIV:1709.02301;%%.

\bibitem{CMSprc:2014ho}
\hrefCMSnoop {}{{CMS Collaboration}, ``{Measurement of higher-order harmonic
  azimuthal anisotropy in PbPb collisions at a nucleon-nucleon center-of-mass
  energy of 2.76 TeV}'',} \textit{ Phys. Rev. C} \textbf{ 89} (2014) 044906,
  \href{http://dx.doi.org/10.1103/PhysRevC.89.044906}{\doi{10.1103/PhysRevC.89.044906}},
\href{http://www.arXiv.org/abs/1310.8651}{\texttt{arXiv:1310.8651}}.
%%CITATION = ARXIV:1310.8651;%%.

\bibitem{STAR:2010}
\hrefCMSnoop {}{{STAR} Collaboration, ``{Charged and strange hadron elliptic
  flow in Cu+Cu collisions at 62.4 and 200 GeV}'',} \textit{ Phys. Rev. C}
  \textbf{ 98} (2010) 044902,
  \href{http://dx.doi.org/10.1103/PhysRevC.81.044902}{\doi{10.1103/PhysRevC.81.044902}},
\href{http://www.arXiv.org/abs/1001.5052}{\texttt{arXiv:1001.5052}}.
%%CITATION = ARXIV:1001.5052;%%.

\bibitem{PHENIX:2007}
\hrefCMSnoop {}{{PHENIX} Collaboration, ``{Scaling properties of azimuthal
  anisotropy in Au+Au and Cu+Cu collisions at $\sqrt{s_{_{\rm NN}}} =
  200$~GeV}'',} \textit{ Phys. Rev. Lett.} \textbf{ 98} (2007) 162301,
  \href{http://dx.doi.org/10.1103/PhysRevLett.98.162301}{\doi{10.1103/PhysRevLett.98.162301}},
\href{http://www.arXiv.org/abs/nucl-ex/0608033}{\texttt{arXiv:nucl-ex/0608033}}.
%%CITATION = NUCL-EX/0608033;%%.

\bibitem{ALICE:2014pPb}
\hrefCMSnoop {}{{ALICE Collaboration}, ``{Multiparticle azimuthal correlations
  in p-Pb and Pb-Pb collisions at the CERN Large Hadron Collider}'',} \textit{
  Phys. Rev. C} \textbf{ 90} (2014) 054901,
  \href{http://dx.doi.org/10.1103/PhysRevC.90.054901}{\doi{10.1103/PhysRevC.90.054901}},
\href{http://www.arXiv.org/abs/1406.2474}{\texttt{arXiv:1406.2474}}.
%%CITATION = ARXIV:1406.2474;%%.

\bibitem{Khachatryan:2015waa}
\hrefCMSnoop {}{{CMS Collaboration}, ``{Evidence for Collective Multiparticle
  Correlations in pPb Collisions}'',} \textit{ Phys. Rev. Lett.} \textbf{ 115}
  (2015) 012301,
  \href{http://dx.doi.org/10.1103/PhysRevLett.115.012301}{\doi{10.1103/PhysRevLett.115.012301}},
\href{http://www.arXiv.org/abs/1502.05382}{\texttt{arXiv:1502.05382}}.
%%CITATION = ARXIV:1502.05382;%%.

\bibitem{CMSplb:2017pp}
\hrefCMSnoop {}{{CMS Collaboration}, ``{Evidence for collectivity in pp
  collisions at the LHC}'',} \textit{ Phys. Lett. B} \textbf{ 765} (2017) 193,
  \href{http://dx.doi.org/10.1016/j.physletb.2016.12.009}{\doi{10.1016/j.physletb.2016.12.009}},
\href{http://www.arXiv.org/abs/1606.06198}{\texttt{arXiv:1606.06198}}.
%%CITATION = ARXIV:1606.06198;%%.

\bibitem{ATLAS:2017pp}
\hrefCMSnoop {}{{ATLAS Collaboration}, ``{Measurement of multi-particle
  azimuthal correlations with the subevent cumulant method in pp and p+Pb
  collisions with the ATLAS detector at the LHC}'',} \textit{ Phys. Rev. C}
  \textbf{ 97} (2018) 024904,
  \href{http://dx.doi.org/10.1103/PhysRevC.97.024904}{\doi{10.1103/PhysRevC.97.024904}},
\href{http://www.arXiv.org/abs/1708.03559}{\texttt{arXiv:1708.03559}}.
%%CITATION = ARXIV:1708.03559;%%.

\bibitem{Ollitrault:1993}
\hrefCMSnoop {}{J.-Y. Ollitrault, ``{Determination of the reaction plane in
  ultrarelativistic nuclear collisions}'',} \textit{ Phys. Rev. D} \textbf{ 48}
  (1993) 1132,
  \href{http://dx.doi.org/10.1103/PhysRevD.48.1132}{\doi{10.1103/PhysRevD.48.1132}},
\href{http://www.arXiv.org/abs/hep-ph/9303247}{\texttt{arXiv:hep-ph/9303247}}.
%%CITATION = HEP-PH/9303247;%%.

\bibitem{Voloshin:1994}
\hrefCMSnoop {}{S.~Voloshin and Y.~Zhang, ``{Flow study in relativistic nuclear
  collisions by Fourier expansion of azimuthal particle distributions}'',}
  \textit{ Z. Phys. C} \textbf{ 70} (1994) 665,
  \href{http://dx.doi.org/10.1007/s002880050141}{\doi{10.1007/s002880050141}},
\href{http://www.arXiv.org/abs/hep-ph/9407282}{\texttt{arXiv:hep-ph/9407282}}.
%%CITATION = HEP-PH/9407282;%%.

\bibitem{Poskanzer:1998yz}
\hrefCMSnoop {}{A.~M. Poskanzer and S.~A. Voloshin, ``{Methods for analyzing
  anisotropic flow in relativistic nuclear collisions}'',} \textit{ Phys. Rev.
  C} \textbf{ 58} (1998) 1671,
  \href{http://dx.doi.org/10.1103/PhysRevC.58.1671}{\doi{10.1103/PhysRevC.58.1671}},
\href{http://www.arXiv.org/abs/nucl-ex/9805001}{\texttt{arXiv:nucl-ex/9805001}}.
%%CITATION = NUCL-EX/9805001;%%.

\bibitem{Alver:2010prc}
\hrefCMSnoop {}{B.~Alver and G.~Roland, ``{Collision geometry fluctuations and
  triangular flow in heavy-ion collisions}'',} \textit{ Phys. Rev. C} \textbf{
  81} (2010) 054905,
  \href{http://dx.doi.org/10.1103/PhysRevC.81.054905}{\doi{10.1103/PhysRevC.81.054905}},
\href{http://www.arXiv.org/abs/1003.0194}{\texttt{arXiv:1003.0194}}.
%%CITATION = ARXIV:1003.0194;%%.

\bibitem{Li:2015v7}
\hrefCMSnoop {}{Y.~Li and J.-Y. Ollitrault, ``{$v_4,v_5,v_6,v_7$: nonlinear
  hydrodynamic response versus LHC data}'',} \textit{ Phys. Lett. B} \textbf{
  744} (2015) 82,
  \href{http://dx.doi.org/10.1016/j.physletb.2015.03.040}{\doi{10.1016/j.physletb.2015.03.040}},
\href{http://www.arXiv.org/abs/1502.02502}{\texttt{arXiv:1502.02502}}.
%%CITATION = ARXIV:1502.02502;%%.

\bibitem{Ollitrault:2009v4}
\hrefCMSnoop {}{J.-Y. Ollitrault, A.~M. Poskanzer, and S.~A. Voloshin,
  ``{Effect of flow fluctuations and nonflow on elliptic flow methods}'',}
  \textit{ Phys. Rev. C} \textbf{ 80} (2009) 014904,
  \href{http://dx.doi.org/10.1103/PhysRevC.80.014904}{\doi{10.1103/PhysRevC.80.014904}},
\href{http://www.arXiv.org/abs/0904.2315}{\texttt{arXiv:0904.2315}}.
%%CITATION = ARXIV:0904.2315;%%.

\bibitem{Yan:2014flc}
\hrefCMSnoop {}{L.~Yan, J.-Y. Ollitrault, and A.~M. Poskanzer, ``{Eccentricity
  distributions in nucleus-nucleus collisions}'',} \textit{ Phys. Rev. C}
  \textbf{ 90} (2014) 024903,
  \href{http://dx.doi.org/10.1103/PhysRevC.90.024903}{\doi{10.1103/PhysRevC.90.024903}},
\href{http://www.arXiv.org/abs/1405.6595}{\texttt{arXiv:1405.6595}}.
%%CITATION = ARXIV:1405.6595;%%.

\bibitem{ALICE:2018xe}
\hrefCMSnoop {}{{ALICE Collaboration}, ``{Anisotropic flow in Xe-Xe collisions
  at $\sqrt{s_{_{\rm NN}}}=5.44$~{TeV}}'',} \textit{ Phys. Lett. B} \textbf{
  784} (2018) 82,
  \href{http://dx.doi.org/10.1016/j.physletb.2018.06.059}{\doi{10.1016/j.physletb.2018.06.059}},
\href{http://www.arXiv.org/abs/1805.01832}{\texttt{arXiv:1805.01832}}.
%%CITATION = ARXIV:1805.01832;%%.

\bibitem{TRK-11-001}
\hrefCMSnoop {}{{CMS Collaboration}, ``{Description and performance of track
  and primary-vertex reconstruction with the CMS tracker}'',} \textit{ JINST}
  \textbf{ 09} (2014) P10009,
  \href{http://dx.doi.org/10.1088/1748-0221/9/10/P10009}{\doi{10.1088/1748-0221/9/10/P10009}},
\href{http://www.arXiv.org/abs/1405.6569}{\texttt{arXiv:1405.6569}}.
%%CITATION = ARXIV:1405.6569;%%.

\bibitem{Chatrchyan:2008zzk}
\hrefCMSnoop {}{{CMS Collaboration}, ``The {CMS} experiment at the {CERN}
  {LHC}'',} \textit{ JINST} \textbf{ 03} (2008) S08004,
  \href{http://dx.doi.org/10.1088/1748-0221/3/08/S08004}{\doi{10.1088/1748-0221/3/08/S08004}}.

\bibitem{GEANT4}
\hrefCMSnoop {}{{GEANT4} Collaboration, ``{Geant4} --- a simulation toolkit'',}
  \textit{ Nucl. Instrum. Meth. A} \textbf{ 506} (2003) 250,
\href{http://dx.doi.org/10.1016/S0168-9002(03)01368-8}{\doi{10.1016/S0168-9002(03)01368-8}}.
%%CITATION = NUIMA,A506,250;%%.

\bibitem{CMS:2017hpt}
\hrefCMSnoop {}{{CMS Collaboration}, ``{Azimuthal anisotropy of charged
  particles with transverse momentum up to 100 GeV/$c$ in PbPb collisions at
  $\sqrt{s_{_{\rm NN}}}=5.02$~{TeV}}'',} \textit{ Phys. Lett. B} \textbf{ 776}
  (2017) 195,
  \href{http://dx.doi.org/10.1016/j.physletb.2017.11.041}{\doi{10.1016/j.physletb.2017.11.041}},
\href{http://www.arXiv.org/abs/1702.00630}{\texttt{arXiv:1702.00630}}.
%%CITATION = ARXIV:1702.00630;%%.

\bibitem{CMSJhep:2011tpc}
\hrefCMSnoop {}{{CMS Collaboration}, ``Long-range and short-range dihadron
  angular correlations in central {PbPb} collisions at $\sqrt{s_{_{\rm
  NN}}}=2.76$~{TeV}'',} \textit{ JHEP} \textbf{ 07} (2011) 076,
  \href{http://dx.doi.org/10.1007/JHEP07(2011)076}{\doi{10.1007/JHEP07(2011)076}},
\href{http://www.arXiv.org/abs/1105.2438}{\texttt{arXiv:1105.2438}}.
%%CITATION = ARXIV:1105.2438;%%.

\bibitem{CMS:2012tpc}
\hrefCMSnoop {}{{CMS Collaboration}, ``Centrality dependence of dihadron
  correlations and azimuthal anisotropy harmonics in {PbPb} collisions at
  $\sqrt{s_{_{\rm NN}}}=2.76$~{TeV}'',} \textit{ Eur. Phys J. C} \textbf{ 72}
  (2012) 10052,
  \href{http://dx.doi.org/10.1140/epjc/s10052-012-2012-3}{\doi{10.1140/epjc/s10052-012-2012-3}},
\href{http://www.arXiv.org/abs/1201.3158}{\texttt{arXiv:1201.3158}}.
%%CITATION = ARXIV:1201.3158;%%.

\bibitem{Sirunyan:2017igb}
\hrefCMSnoop {}{{CMS Collaboration}, ``{Pseudorapidity and transverse momentum
  dependence of flow harmonics in pPb and PbPb collisions}'',} \textit{ Phys.
  Rev. C} \textbf{ 98} (2018) 044902,
  \href{http://dx.doi.org/10.1103/PhysRevC.98.044902}{\doi{10.1103/PhysRevC.98.044902}},
\href{http://www.arXiv.org/abs/1710.07864}{\texttt{arXiv:1710.07864}}.
%%CITATION = ARXIV:1710.07864;%%.

\bibitem{Luzum:2012da}
\hrefCMSnoop {}{M.~Luzum and J.-Y. Ollitrault, ``{Eliminating experimental bias
  in anisotropic-flow measurements of high-energy nuclear collisions}'',}
  \textit{ Phys. Rev. C} \textbf{ 87} (2013) 044907,
  \href{http://dx.doi.org/10.1103/PhysRevC.87.044907}{\doi{10.1103/PhysRevC.87.044907}},
\href{http://www.arXiv.org/abs/1209.2323}{\texttt{arXiv:1209.2323}}.
%%CITATION = ARXIV:1209.2323;%%.

\bibitem{Bilandzic:2013kga}
A.~Bilandzic\hrefCMSnoop {}{ {et~al.}, ``{Generic framework for anisotropic
  flow analyses with multiparticle azimuthal correlations}'',} \textit{ Phys.
  Rev. C} \textbf{ 89} (2014) 064904,
  \href{http://dx.doi.org/10.1103/PhysRevC.89.064904}{\doi{10.1103/PhysRevC.89.064904}},
\href{http://www.arXiv.org/abs/1312.3572}{\texttt{arXiv:1312.3572}}.
%%CITATION = ARXIV:1312.3572;%%.

\bibitem{Lokhtin:2006hmc}
\hrefCMSnoop {}{I.~P. Lokhtin and A.~M. Snigirev, ``{A model of jet quenching
  in ultrarelativistic heavy ion collisions and high-$p_{\rm T}$ hadron spectra
  at RHIC}'',} \textit{ Eur. Phys. J.} \textbf{ 45} (2006) 211,
  \href{http://dx.doi.org/10.1140/epjc/s2005-02426-3}{\doi{10.1140/epjc/s2005-02426-3}},
\href{http://www.arXiv.org/abs/hep-ph/0506189}{\texttt{arXiv:hep-ph/0506189}}.
%%CITATION = ARXIV:hep-ph/0506189;%%.

\bibitem{IP:Glasma}
\hrefCMSnoop {}{B.~Schenke and S.~Schlichting, ``{3-D Glasma initial state for
  relativistic heavy ion collisions}'',} \textit{ Phys. Rev. C} \textbf{ 94}
  (2016) 044907,
  \href{http://dx.doi.org/10.1103/PhysRevC.94.044907}{\doi{10.1103/PhysRevC.94.044907}},
\href{http://www.arXiv.org/abs/1605.07158}{\texttt{arXiv:1605.07158}}.
%%CITATION = ARXIV:1605.07158;%%.

\bibitem{Schenke:2010ms}
\hrefCMSnoop {}{B.~Schenke, S.~Jeon, and C.~Gale, ``{3+1D hydrodynamic
  simulation of relativistic heavy-ion collisions}'',} \textit{ Phys. Rev. C}
  \textbf{ 82} (2010) 014903,
  \href{http://dx.doi.org/10.1103/PhysRevC.82.014903}{\doi{10.1103/PhysRevC.82.014903}},
\href{http://www.arXiv.org/abs/1004.1408}{\texttt{arXiv:1004.1408}}.
%%CITATION = ARXIV:1004.1408;%%.

\bibitem{Ryu:2015bv}
S.~Ryu\hrefCMSnoop {}{ {et~al.}, ``{Importance of the bulk viscosity of QCD in
  ultrarelativistic heavy-ion collisions}'',} \textit{ Phys. Rev. Lett.}
  \textbf{ 115} (2015) 132301,
  \href{http://dx.doi.org/10.1103/PhysRevLett.115.132301}{\doi{10.1103/PhysRevLett.115.132301}},
\href{http://www.arXiv.org/abs/1502.01675}{\texttt{arXiv:1502.01675}}.
%%CITATION = ARXIV:1502.01675;%%.

\bibitem{UrQMD:2008}
H.~Petersen\hrefCMSnoop {}{ {et~al.}, ``{Fully integrated transport approach to
  heavy ion reactions with an intermediate hydrodynamic stage}'',} \textit{
  Phys. Rev. C} \textbf{ 78} (2008) 044901,
  \href{http://dx.doi.org/10.1103/PhysRevC.78.044901}{\doi{10.1103/PhysRevC.78.044901}},
\href{http://www.arXiv.org/abs/0806.1695}{\texttt{arXiv:0806.1695}}.
%%CITATION = ARXIV:0806.1695;%%.

\bibitem{Giacalone:2017}
\hrefCMSnoop {}{G.~Giacalone, J.~Noronha-Hostler, M.~Luzum, and J.-Y.
  Ollitrault, ``{Hydrodynamic predictions for 5.44 TeV Xe+Xe collisions}'',}
  \textit{ Phys. Rev. C} \textbf{ 97} (2018) 034904,
  \href{http://dx.doi.org/10.1103/PhysRevC.97.034904}{\doi{10.1103/PhysRevC.97.034904}},
\href{http://www.arXiv.org/abs/1711.08499}{\texttt{arXiv:1711.08499}}.
%%CITATION = ARXIV:1711.08499;%%.

\bibitem{Moreland:2014tr}
\hrefCMSnoop {}{J.~S. Moreland, J.~E. Bernhard, and S.~A. Bass, ``{Alternative
  ansatz to wounded nucleon and binary collision scaling in high-energy nuclear
  collisions}'',} \textit{ Phys. Rev. C} \textbf{ 92} (2015) 011901(R),
  \href{http://dx.doi.org/10.1103/PhysRevC.92.011901}{\doi{10.1103/PhysRevC.92.011901}},
\href{http://www.arXiv.org/abs/1412.4708}{\texttt{arXiv:1412.4708}}.
%%CITATION = ARXIV:1412.4708;%%.

\bibitem{Xe:Shape}
\hrefCMSnoop {}{P.~Moller, S.~A. J., T.~Ichikawa, and H.~Sagawa, ``{Nuclear
  ground-state masses and deformations: FRDM(2012)}'',} \textit{ Atom. Data
  Nucl. Data Tabl.} \textbf{ 109-110} (2015) 1,
  \href{http://dx.doi.org/10.1016/j.adt.2015.10.002}{\doi{10.1016/j.adt.2015.10.002}},
\href{http://www.arXiv.org/abs/1508.06294}{\texttt{arXiv:1508.06294}}.
%%CITATION = ARXIV:1508.06294;%%.

\bibitem{Giacalone:2017a}
\hrefCMSnoop {}{G.~Giacalone, J.~Noronha-Hostler, and J.-Y. Ollitrault,
  ``{Relative flow fluctuations as a probe of initial state fluctuations}'',}
  \textit{ Phys. Rev. C} \textbf{ 95} (2017) 054910,
  \href{http://dx.doi.org/10.1103/PhysRevC.95.054910}{\doi{10.1103/PhysRevC.95.054910}},
\href{http://www.arXiv.org/abs/1702.01730}{\texttt{arXiv:1702.01730}}.
%%CITATION = ARXIV:1702.01730;%%.

\bibitem{size:fluc}
\hrefCMSnoop {}{R.~S. Bhalerao, M.~Luzum, and J.-Y. Ollitrault,
  ``{Understanding anisotropy generated by fluctuations in heavy-ion
  collisions}'',} \textit{ Phys. Rev. C} \textbf{ 84} (2011) 054901,
  \href{http://dx.doi.org/10.1103/PhysRevC.84.054901}{\doi{10.1103/PhysRevC.84.054901}},
\href{http://www.arXiv.org/abs/1107.5485}{\texttt{arXiv:1107.5485}}.
%%CITATION = ARXIV:1107.5485;%%.

\bibitem{size:visc}
\hrefCMSnoop {}{P.~Romatschke and U.~Romatschke, ``{Viscosity information from
  relativistic nuclear collisions: how perfect is the fluid observed at
  RHIC?}'',} \textit{ Phys. Rev. Lett.} \textbf{ 99} (2007) 172301,
  \href{http://dx.doi.org/10.1103/PhysRevLett.99.172301}{\doi{10.1103/PhysRevLett.99.172301}},
\href{http://www.arXiv.org/abs/0706.1522}{\texttt{arXiv:0706.1522}}.
%%CITATION = ARXIV:0706.1522;%%.

\end{thebibliography}\endgroup
